# The geometry of flow: Advancing predictions of river geometry with multi-model machine learning

**Authors:** Shuyu Y Chang[1,2], Zahra Ghahremani[3], Laura Manuel[4], Mohammad Erfani[5], Chaopeng Shen[6], Sagy Cohen[7], Kimberly Van Meter[1,2], Jennifer L Pierce[3], Ehab A Meselhe[4], Erfan Goharian[5]

[1]*Department of Geography, Pennsylvania State University, 302 Walker Building, University Park, PA 16803*

[2]*Earth and Environmental Systems Institute, 2217 Earth-Engineering Sciences Building*

*University Park, PA 16802-6813*

[3]*Department of Geoscience, Boise State University, Environmental Research Building 1160, Boise, ID 83725*

[4]*Department of River-Coastal Science and Engineering, Tulane University, 6823 St. Charles Avenue*

*New Orleans, LA 70118*

[5]*Department of Civil and Environmental Engineering, Univerity of South Carolina, 300 Main St, Room C206, Columbia, SC 29208*

[6]*Department of Civil and Environmental Engineering Engineering, Pennsylvania State University, 231C Sackett Bulding, University Park, PA 16803*

[7]*Department of Geography, University of Alabama, Shelby Hall 2019-E, Tuscaloosa, AL, 35487*

**Corresponding authors' Email:** cshen@engr.psu.edu

sagy.cohen@ua.edu

vanmeterkvm@psu.edu

**Abstract**

Hydraulic geometry parameters describing river hydrogeomorphic is important for flood forecasting. Although well-established, power-law hydraulic geometry curves have been widely used to understand riverine systems and mapping flooding inundation worldwide for the past 70 years, we have become increasingly aware of the limitations of these approaches. In the present study, we have moved beyond these traditional power-law relationships for river geometry, testing the ability of machine-learning models to provide improved predictions of river width and depth. For this work, we have used an unprecedentedly large river measurement dataset (HYDRoSWOT) as well as a suite of watershed predictor data to develop novel data-driven approaches to better estimate river geometries over the contiguous United States (CONUS). Our Random Forest, XGBoost, and neural network models out-performed the traditional, regionalized power law-based hydraulic geometry equations for both width and depth, providing R-squared values of as high as 0.75 for width and as high as 0.67 for depth, compared with R-squared values of 0.57 for width and 0.18 for depth from the regional hydraulic geometry equations. Our results also show diverse performance outcomes across stream orders and geographical regions for the different machine-learning models, demonstrating the value of using multi-model approaches to maximize the predictability of river geometry. The developed models have been used to create the newly publicly available STREAM-geo dataset, which provides river width, depth, width/depth ratio, and river and stream surface area (%RSSA) for nearly 2.7 million NHDPlus stream reaches across the rivers and streams across the contiguous US.

**Plain Language Summary**

Scientists and river managers use measurements of river geometry such as width and depth to forecast floods and understand river behavior. However, the methods used to estimate river geometry that have been used for decades are imprecise and thus lead to poor predictions of river discharge dynamics. Here, we've used new machine learning-based modeling approaches to provide better predictions of river width and depth. We tested different machine-learning models, which were developed based on the HYDRoSWOT set of measurements of rivers across the U.S. These new models all provide better estimates of river width and depth than the old methods. Our research can help us to provide better estimates of flood dynamics and improve our understanding of rivers across the U.S.

**Main Points**

1. Machine Learning models outperform regional (physiographic) hydraulic geometry equations for predicting stream width and depth.
2. Model performance varies by stream order and geographical region, demonstrating the utility of multi-model machine-learning approaches.

3. The STREAM-Geo river geometry dataset based on our models provides predictions of river width, depth, width-to-depth ratio, and river surface area for all NHDPlus stream reaches.

# 1.0 Introduction

Rivers and streams transport water, sediments, and nutrients along the channel and between the channel and floodplain, from inland to coastal regions, providing essential resources, supporting biodiversity and ecological functions, and shaping landscapes. Rivers are inherently dynamic, constantly changing in response to changes in both seasonal and event-based changes in precipitation and runoff (Rhoads, 2020). With changes in land use and climate, river dynamics are also changing over time, with an increased incidence of flooding in many regions (Kundzewicz et al., 2014). Globally, it has been estimated that the current 100-year flood may occur at least twice as frequently across 40% of the earth's surface by the year 2050 (Arnell & Gosling, 2016). In the U.S., it is estimated that approximately 41 million Americans currently live within the 100-year floodplain, and it is expected that exposure to flood risks will continue to increase as a function of both climate and land-use change (Wing et al., 2018). With these changes, it is a priority to provide better predictions of long-term flood risk as well as better event-based forecasting in order to ensure public safety and to meet water-resource needs (Adams, 2016). In the 1970s and 1980s, soon after the establishment of the 13 U.S. River Forecast Centers, flood forecasting was primarily based on locally developed, event-based hydrologic models (Adams, 2016; NOAA, 2018). Today, however, the U.S. National Weather Service (NWS) relies heavily on the National Water Model, which provides continuous estimates of streamflow and other hydrologic indicators (Cosgrove et al., 2019). The National Water Model includes a network of 2.7 million river reaches, coupled with 1-km grid scale land surface modeling and 250-m grid scale modeling of surface and subsurface runoff (Cosgrove et al. 2019).

To support large-scale modeling efforts such as those of the National Water Model, it is crucial to accurately represent the geomorphological features of the stream network. In particular, widths and depths of rivers are fundamental characteristics of river geometry and have profound effects on flood magnitudes, frequency, and inundation (Cohen et al., 2014; Knight & Demetriou, 1983; Knox, 1993; Magilligan, 1992; Neal et al., 2015) as well as hydraulic habitat conditions (Choi et al., 2018; Kingma & Ba, 2017; Lamouroux et al., 1995), stream biological communities (Rhoads et al., 2003; Stewardson, 2005), and downstream water quality (Baradei, 2020; Han et al., 2019; Liu et al., 2021; NOVO et al., 1989).

Due to the natural complexity of channels, the costliness of bathymetric observations, and the high degree of computational power needed to process big data, large-scale hydrologic watershed models often have to simplify their channel cross-section representations and thus rely heavily on simple empirical relationships for river geometry parameter estimations. For example, the National Water Model assumes trapezoidal-shaped channels with 11 parameters, including the top width, bottom width, side slope, and Manning's n, to forecast streamflow for the 2.7 million U.S. river reaches. In addition, regional (physiographic) hydraulic curves, similar to hydraulic geometry curves (Leopold & Maddock Jr., 1953), which demonstrate a power-law relationship between the catchment drainage area and the bankfull characteristics, are used to determine the bankfull width and cross-sectional area in the National Water Model .

The power-law-based hydraulic geometry curves describing relationships between streamflow and river mean depth and width were first proposed by Leopold and Maddock (Leopold & Maddock Jr., 1953), as follows:

$$w = aQ^b \tag{1}$$

$$d = cQ^f \tag{2}$$

where Q is discharge, *w* is the mean width and *d* is the mean depth at a channel cross-section. In these equations, *a, b, c,* and *f* are coefficient parameters obtained via log-log linear regression analysis based on survey data. In practice, these equations are usually simplified further, relying on catchment area instead of discharge in order to allow for prediction in ungauged basins. Unfortunately, this regional regression approach suffers from substantial variabilities and uncertainties (Gleason, 2015; Harman et al., 2008; Knighton, 1975; Phillips, 1990; Stewardson, 2005), mainly due to either the poor quality or quantity of hydraulic observations (lack of data support), the natural spatial and temporal variability of rivers (not considering factors other than the drainage area), and theoretical shortcomings (lack of theoretical understanding of how various hydraulic and geometric factors determine hydraulic geometry). In 1973, Richards (1973) challenged the two-parameter power-law model for describing river geometry by providing an alternative solution—a polynomial form. More recently, Gleason (2015) has called for new advances in understanding hydraulic geometry and the next generation of river geometry models.

Indeed, current research clearly shows that incorporating more realistic channel geometry could better capture river dynamics and thus substantially improve the performance of hydrologic models. For example, Brackins et al., (2021) suggest the application of two newly proposed river geometry representations (one is the four points thalweg and the other is the five evenly spaces points), demonstrating that such an approach reduced peak errors by 80%–93% and reducing the stage root mean square error (RMSE) by 14%-54% in the Hydrologic Engineering Center's River Analysis System (HEC-RAS) compared with use of the simple trapezoidal shape in the National Water Model v1.2, at three test sites. In addition, Heldmyer et al., (2021) demonstrated that using updated channel parameters led to improvements in streamflow simulations for 7,400 gauge locations, with a statistically significant mean $R^2$ increase from 0.479 to 0.494.

Recent advances in remote sensing, especially with the launch of the Surface Water and Ocean Topography (SWOT) mission in 2022, will also provide new opportunities for the hydrogeomorphology community to address these persisting problems to represent river geometry at large scales. For example, Durand et al., (2008) have developed a novel data-assimilation-based approach to characterize channel bathymetry using synthetic SWOT water surface elevation measurements, with their results showing substantial improvements in depth and slope estimations. Despite the promise of this approach, it is important to note here that SWOT can only characterize rivers wider than 100 m, and that it provides only limited temporal sampling (Durand et al., 2010; Frasson et al., 2019). Therefore, the two main limitations of SWOT-derived surface water products must be carefully considered: (1) a lack of detailed information and measurements for headwater and other small streams; and (2) uneven temporal sampling, leading to a limited ability to observe long-term river geometry dynamics and flood events.

Machine learning provides another promising direction for improving representations of river geometry. There have been two important pioneer studies examining the efficacy of machine learning regression models to estimate cross-sectional river geometry. Lin et al., (2020) developed an XGBoost approach, surveying 16 environmental covariates to estimate the bankfull river width, which provided considerable predictive power, with an $R^2 = 0.81$ for a global-scale test dataset and $R^2 = 0.77$ for North American test cases. These results contrast with the power-law parameterization schemes, which only capture 30–40% of the variance in river width (Lin et al., 2020). This study, however, was limited only to width predictions. For depth, Raney et al. (2021) developed a calibrated Multilayer Perceptron (MLP) model, increasing the $R^2$ from 0.32 with regional power law-based predictions to 0.58 for mean river depth predictions based on 3,695 stream reaches in the United States (Raney et al., 2021). These results suggest that there are great opportunities for machine-learning approaches to propel advances in hydrogeomorphology.

In this paper we apply data-driven approaches to create a framework for determining channel geometry (i.e., width and depth), using publicly available catchment and channel characteristics datasets, and an unprecedentedly large continental-scale river geometry observation database. In this work, we are attempting to answer the following questions: (1) How do machine learning-based models compare to traditional power-law approaches in their ability to predict stream width and depth? (2) How does model performance vary across stream types and geographical regions? Answering these research questions allows us to work towards two broader objectives. The first is to evaluate the performance of data-driven approaches to estimate river geometry and thus pave the way for improving theoretical knowledge and developing end-to-end models customized for many continental-scale hydrology models. The second is to develop a comprehensive, high-quality, fine-scale river geometry database across the contiguous US, STREAM-geo. STREAM-geo is a free and publicly available dataset of river width, river depth, river width to depth ratio, and the river and stream surface area (RSSA) for the nearly 2.7 million reaches across the contiguous US, based on NHDPLUS flowlines.

## 2.0 Data and Methods

## 2.1 Data Sources and Preparation

### 2.1.1 Stream Geometry Data

In-situ (at-a-station) hydraulic and channel geometry observations utilized in this work were obtained from the HYDRoacoustic dataset in support of Surface Water Oceanographic Topography (HYDRoSWOT), published through the United States Geological Survey (USGS) ScienceBase (Canova et al., 2016). The current HYDRoSWOT dataset includes 220,000 + records of hydraulic characteristics (mean depth, mean velocity, discharge, stage, water-surface width, maximum depth, maximum velocity, etc.) of river cross sections for more than 5,000 stations and corresponding stream gauge metadata over the United States. These data were collected through the use of Acoustic Doppler Current Profilers (ADCPs) by USGS teams from the 1940s through 2014. The dataset represents stream hydraulic characteristics under different flow conditions, ranging from smaller headwaters running in narrow deep valleys with steep banks to the widest river in the U.S., the Mississippi River.

The water-surface width and weighted mean river depth (**Figure 1a**) were used as target variables for subsequent modeling. To exclude unrealistic values, outliers (width and depth values above the 95th percentile or below the 5th percentile for each site) were removed from the dataset. All no-data (NaN) values were also removed. **Figure 1**, panels **(b)-(e)**, show the locations and distributions of width and depth measurements from the USGS ADCP HYDRoSWOT dataset. Note that, overall, there are more river width data points than river depth data points, and also that the eastern U.S. has more intensive sampling than the west, leading to an uneven distribution of in-situ river geometry knowledge.

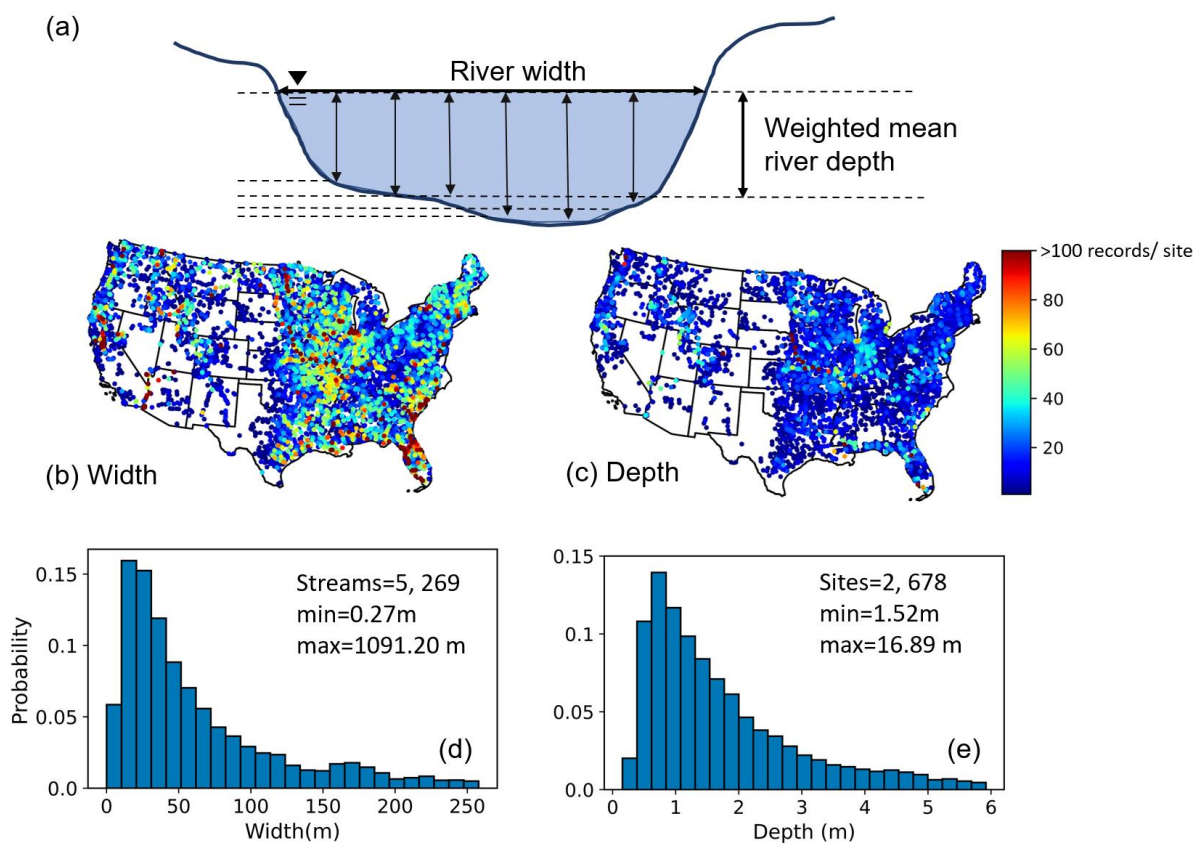

**Figure 1.** Observed river depth and width data across the contiguous US. (a) Schematic of river width and depth definitions, as obtained from HYDRoSWOT; (b) and (c): the spatial distribution of measured sites and the number of records per site for river width and depth; (d) and (e): the histograms of river width and depth.

After data cleaning, for sites with multiple observations, we calculated the median width and depth values from the time series data for sites with more than 5 records; sites with fewer than 5 data points were removed from the dataset. This provided a dataset with time-stationary values for each site (single observation) and thus allowed for the prediction of a median stream width and depth. The finalized training dataset contained width measurements for 5,269 stream reaches and depth measurements for 2,678 stream reaches. Hereafter we refer to these values as median width and median depth. Summary statistics for the final HYDRoSWOT input dataset are provided in the Supplemental Information (**SI Table S1**).

## 2.1.2 Stream and Catchment Data

A total of 17 stream and catchment attributes were collected from numerous datasets for use in model development. These datasets include the National Hydrography Dataset Plus (NHDPlus) Version 2.0 and the Select Attributes for NHDPlus Version 2.1 (Schwarz et al., 2018), the Gridded Soil Survey Geographic Database (gSSURGO) (Natural Resources Conservation Service, 2016), and the Global Aridity Index and Potential Evapotranspiration (ET0) Climate Database v3 (Zomer et al., 2022). Variables utilized from these datasets include climatic variables such as the aridity index ($AI = \frac{P}{PET}$) (Zomer et al., 2022); hydrological variables, including stream flow, extracted from (Canova et al., 2016); reach metadata such as elevation, slope, stream order (NHDPlus Version 2.0); river bedding variables such as D50 (Abeshu et al., 2022); upstream catchment surface characteristics such as catchment area (NHDPlus Version 2.0), land use (Schwarz et al., 2018), and seasonal vegetation indexes (Schwarz et al., 2018); subsurface variables such as soil types (Natural Resources Conservation Service, 2016); and anthropogenic features such as Population density per catchment and accumulated dam construction (Schwarz et al., 2018).

Using the HYDRoSWOT point locations as a base dataset, the locations were matched with the closest NHDPlus reach within a 500-m buffer, and NHDPlus attributes for those reaches were then associated with the individual measured locations. Soil properties and aridity index were extracted by averaging the raster values within a 500-m buffer. For a summary of all predictor datasets as well as relevant data sources and notes on processing, see **SI Table S2**. Supplemental **SI Figure S1** provides maps of eight selected attributes for all the HYDRoSWOT gauged streams.

## 2.2 River Geometry Model Development

In the current study, we used four different approaches to predict river geometry (**Figure 2**). In the first approach, we used the simple, regionally varying power-law approach that is currently utilized in the National Water Model (Blackburn-Lynch et al., 2017). We then employed three additional data-driven modeling approaches: (1) multiple linear regression, (2) tree-based approaches (Random Forest, XGBoost), and (3) a neural network (multilayer perceptron).

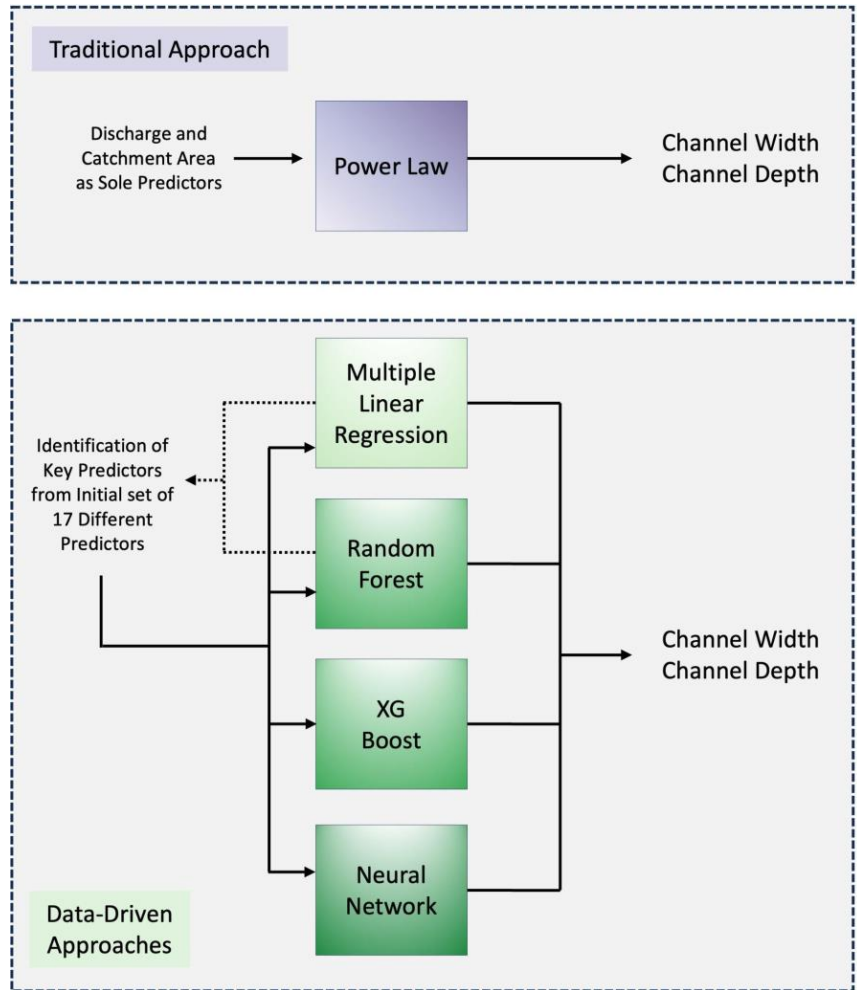

**Figure 2.** Modeling approaches to predicting stream channel width and depth. Note that while traditional approaches use only watershed area and streamflow (discharge) to predict channel width and depth, our data-driven, machine-learning approaches use 17 different predictors.

### 2.2.1 Power Law

As a comparison, regional power-law relationships were developed to predict channel width and depth:

$$w = aA_w{}^b \tag{3}$$

$$d = cA_w{}^d \tag{4}$$

where $A_w$ is the catchment area, $w$ is the river width, and $d$ the river depth at a channel cross-section, and *a, b, c,* and *f* are regression coefficients. Regional curves are typically developed for a single physiographic province with similar landforms (Bieger et al., 2015). Accordingly, in the present study, we developed 20 different equations for the 20 Hydrologic Landscape Regions (Wolock et al., 2004) over the CONUS.

### 2.2.2 Linear regression

Both simple and multiple linear regression models were used in this study. Simple linear regression analysis was carried out to more explicitly identify how watershed characteristics and other relevant predictors co-vary with stream width and depth. For this portion of the analysis, linear regression equations were developed, slope values were obtained, and the strength of the correlation was evaluated based on $R^2$-values and p-values (Simple linear regression for river width: **SI Figure S4**, **SI Table 9**; simple linear regression for river depth: **SI Figure S5**, **SI Table 10**).

Multiple linear regression was carried out using JMP statistical software (SAS Instittute Inc., 2023) and was optimized for each target-dependent variable (median channel width and depth). The median width and median depth datasets were used in the analysis, and all variables were log-transformed, given the high degree of skewness in most variables. All variables presented in **SI Table S2** were used to initiate the model. A forward stepwise regression technique was then used to determine the significance of each variable to the model performance. Variables were initially removed based on p-value criteria, with p-values >0.05 being considered not significant. Next, variables were removed in a stepwise fashion if their removal did not reduce the R-squared value by more than 0.05. For the final MLR equations, see **SI SE1** and **SI SE2**.

### 2.2.3 Decision Tree-based models

Random forest regression (RFR) and extreme gradient boosting (XGBoost), two popular decision tree approaches, were applied in this study. Random forest, a supervised learning algorithm, uses a bagging technique to build many decision trees and thus reduce the likelihood of overfitting. Its ability to provide the feature importance of each variable and to handle multicollinearity, missing data, high-dimensional data, and nonlinear relationships have led to its increasing use in modeling environmental systems (Ziegler and König, 2013). The XGBoost algorithm is another tree-based model that typically outperforms the random forest model and utilizes boosting and gradient boosting to improve the model by resolving the weakness of the previous model (Chollet, 2018).

The GridSearchCV method (scikit-learn Python) (Pedregosa et al., 2018) was applied to optimize the following hyperparameters for the random forest: maximum tree depth, number of trees, number of features to check for the best split, the minimum samples at a leaf node, and the minimum samples to split an internal node. Details regarding the hyperparameter space are provided in **SI Table S3** for random forest and **SI Table S4** for XGBoost**.** Five-fold cross-validation (scikit-learn Python) (Pedregosa et al., 2018) was also used with the training dataset to prevent overfitting.

#### 2.2.4 Multilayer perceptron (MLP) model

Deep neural networks can harness the power of big data, capturing complex data distributions, detecting unrecognized linkages, turning raw data into readily useful information and even facilitating the development of physically based models (Shen, 2018; Shen et al., 2023). In the field of hydrogeomorphology, they have been used to study changes in river delta morphology (Munasinghe et al., 2021), to classify terrestrial drainage networks (Donadio et al., 2021), to extract large-scale riverine bathymetry (Ghorbanidehno et al., 2021), to detect riverine bridges (Chen et al., 2021), etc.

In the current study, we employed a multilayer perceptron (MLP), one of the most popular neural network architectures (Hornik et al., 1989). The architecture of MLPs can be described as a set of connections from an input layer, through one or more hidden layers, to an output layer. Each hidden layer contains one or more neurons, where each neuron, with a set of trainable weights and a bias, computes a weighted sum of inputs and then passes the result through an activation function (adding non-linearity to the model) (Abiodun et al., 2018).

Data scaling and data transformation are critical steps in the development of deep learning models such as the multi-layer perceptron, as input variables may be drawn from different domains and, in turn, may have different scales, units, and distributions. For this purpose, we applied power-law transformations to three variables (discharge, median particle size (D50), and slope), and the remaining variables were rescaled to a Standard Gaussian function with a mean of 0 and a unit standard deviation. Min-max normalization was then applied to the standardized variables. For additional details regarding scaling methods, see **SI Table S6**.

The MLP parameter and hyperparameter search spaces were sampled to approach the optimal network topology and hyperparameter space to best estimate river geometry. Details regarding the hyperparameter space are provided in **SI Table S5**. Overall, 300 hyperparameter sets were generated by randomly sampling hyperparameter values from the established ranges. To find the optimal hyperparameter combination, we trained the model 300 times with an early stop function (a technique used to prevent overfitting) based on the training dataset with 5-fold cross-validation (scikit-learn Python). The state-of-the-art hyperparameter space optimization algorithm, the Adam stochastic gradient descent algorithm (Kingma & Ba, 2017), was used to optimize a Mean Squared Error (MSE) loss function. The hyperparameter candidates of the top-performing model based on the validation dataset (20% of the training dataset) were selected for the final round of model fine-training.

#### 2.3 Performance Indicators and Model Analysis

For model training and performance evaluation, the input dataset was shuffled and randomly distributed to a 75:25 split of training and testing datasets. The evaluation metrics used to evaluate model performance were the Root Mean Squared Error (RMSE), Percent Bias (PBIAS), coefficient of determination ($R^2$), and Nash Sutcliffe efficiency ($NSE$).

From the Random Forest model, the predictor importance was quantified using the method of permutation-based mean-squared error (MSE) reduction (Grömping 2009). Using this approach, we calculated the out-of-bag MSE (OOBMSE) for each tree as the average of the squared deviations of out-of-bag responses from their respective predictions (Van Meter 2020). If the model performance has little or no predictive value for the response, there will be no appreciable difference in model results due to the permutation of that predictor. Accordingly, for each variable in each tree, the difference between the permuted and non-permuted OOBMSE is calculated, as is the average change in MSE across all trees.

## 3.0    Results and Discussion

## 3.1    Key Predictors of River Width and Depth

The traditional power-law approach to river geometry relies on discharge or area alone to predict river width and depth **(Eq. 1-4)**. In the National Water Model, the power law-based predictions are augmented by an empirical, region-specific parameterization due to the important effects of climate, land use, and land cover on these non-linear relationships between river geometry and discharge.

In contrast, in our data-driven approach, we have used a total of 17 predictors to move beyond just discharge and area as a means of improving our predictions of river geometry. These predictors include metrics associated with geology, climate, land use, and other anthropogenic factors (**SI Table S2**). To explore the predictor importance, we first carried out a simple regression analysis with the objective of characterizing the strength and the nature of the individual correlation relationships (**SI Figure. S4** and **Table S9** for river width; **SI Figure. S5** and **Table S10** for river depth). Not surprisingly, for both width and depth, the linear regression analysis showed that discharge explains the majority of the variability in median width ($R^2 = 0.61$) and depth ($R^2 = 0.57$), with catchment area alone also providing explanatory power (width, $R^2 = 0.36$; depth, $R^2 = 0.25$). Stream order also has strong explanatory power, which is to be expected due to the strong correlation between stream order and catchment area (Downing, 2010; Lin et al., 2020). Another important predictor of both width and depth is the number of upstream dams, with more dams being associated with wider and deeper streams (width, $R^2 = 0.21$; depth, $R^2 = 0.16$). In contrast, there is a negative relationship between watershed vegetation and stream width and depth, with higher enhanced vegetation index (EVI) values being associated with smaller widths and depths.

Next, we extracted relative feature importance (RFI) results from the random forest model to further our understanding of the relevance and contribution of each predictor variable in predicting width and depth (**Figure 3**). Again, the random forest results show that for both width and depth, the median river discharge, $Q$, is by far the most important predictor ($RFI_{width} = 0.67$, $RFI_{depth} = 0.78$), with catchment area being the second-most important predictor for width. The outsized impacts of annual median Q suggest that this metric brings in unique information, information that cannot be provided by other attributes. This information might include missing climatic features, e.g., extreme rainfall frequency and magnitudes, higher-moment topographic patterns, e.g., converging vs. diverging landscapes, or surface properties, e.g., the infiltrating capacity of upstream soils. Beyond these top predictors, we see some variability in predictor importance between the width and the depth models. For width, vegetation and aridity are the next-most important predictors, while for depth, the elevation of the stream reach and catchment slope are more important (**Figure 3**). Interestingly, while the number of dams appears to have a great deal of explanatory power for width and depth in the linear regression analysis, as discussed above, the importance of this predictor is less than that of other geomorphological and land use-related variables in the random forest ($RFI_{width} = 0.01$, $RFI_{depth} = 0.01$).

Elevation and slope are the second-tier predictors after discharge (**Figure 3**), suggesting that mountainous streams are characteristically different from those on the plains. Presumably, this is related to the higher stream power (and thus incision) and tougher substrates in the mountains. Field studies and data analysis in (Whitehead et al. 2009) have shown bedrock channels to be narrower and deeper than alluvial channels. However, bedrock datasets are not available at the national scale, and we suspect slope and elevation act as a proxy for such information.

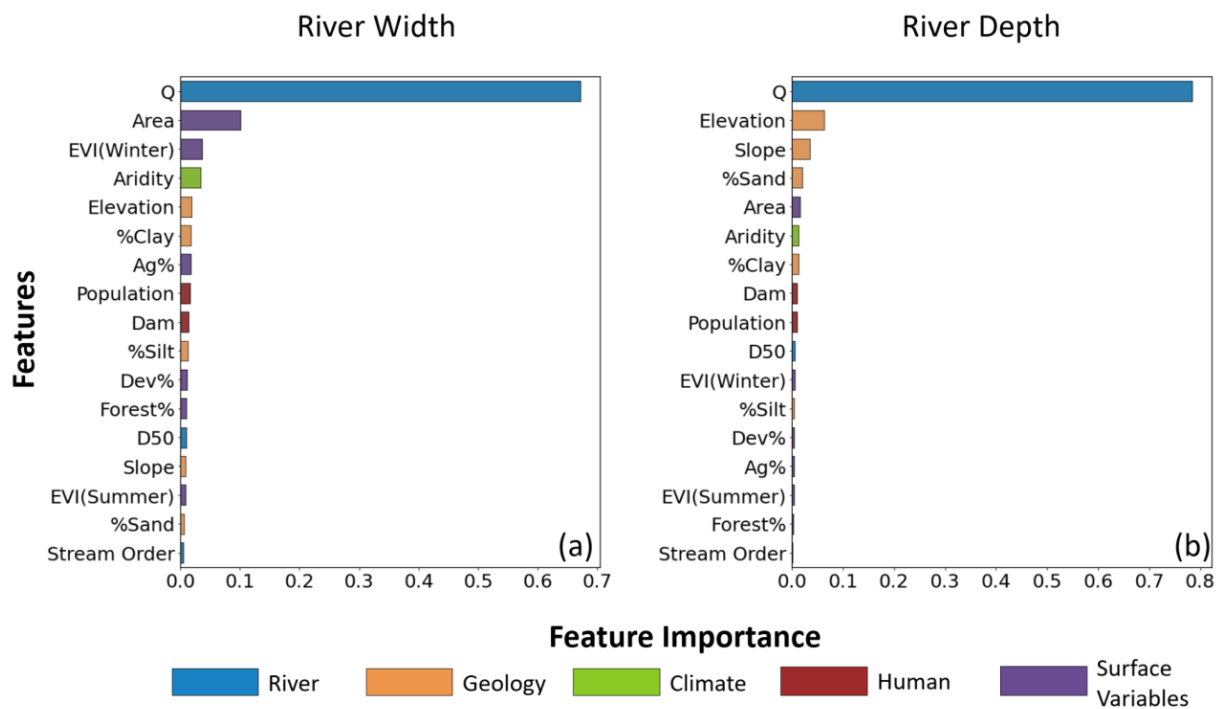

**Figure 3.** Relative predictor importance of variables used in the machine-learning models, as determined from the random forest model. Predictor importance is shown for both (**a**) median river width and (**b**) median river depth.

### 3.2 Overall Model Performance

The random forest, XGBoost, and neural network (MLP) models all performed well in predicting river width and depth under median-flow conditions, providing more accurate results than the traditional, regionalized power-law methodology used for prediction in the U.S. National Water Model and other large-scale models (**Figure 4**). For the median river width, the random forest model performed the best (NSE$_{RF}$ = 0.74, R$^2_{RF}$ = 0.75; NSE$_{XGB}$ = 0.73, R$^2_{XGB}$ = 0.73; NSE$_{MLP}$ = 0.73, R$^2_{MLP}$ = 0.73), although the performance was comparable for all three models. In contrast, the national-scale multiple linear regression model performed even worse than the power-law predictions, explaining only 33% of the variance in the median river width, compared with the 45% explained by the power law (R$^2_{MLR}$ = 0.33, NSE$_{MLR}$ = -0.26). For the median river depth, the accuracy of the machine-learning models was not as good as that for width (NSE$_{RF}$ = 0.63, R$^2_{RF}$ = 0.67; NSE$_{XGB}$ = 0.64, R$^2_{XGB}$ = 0.67; NSE$_{MLP}$ = 0.57, R$^2_{MLP}$ = 0.66), but, again, all three of the machine-learning models performed similarly. The multiple linear regression model performed better for depth than for width (R$^2_{MLR}$ = 0.66, NSE$_{MLR}$=0.52), but the performance is still poorer than that of the machine-learning models. Due to the consistently poor results for the MLR model, we focus only on comparing the tree-based and neural network model results with the traditional power-law approach in the rest of the paper. For additional results regarding model performance metrics for median river width and depth, see **SI Table 11** for river width, and **SI Table 12** for river depth.

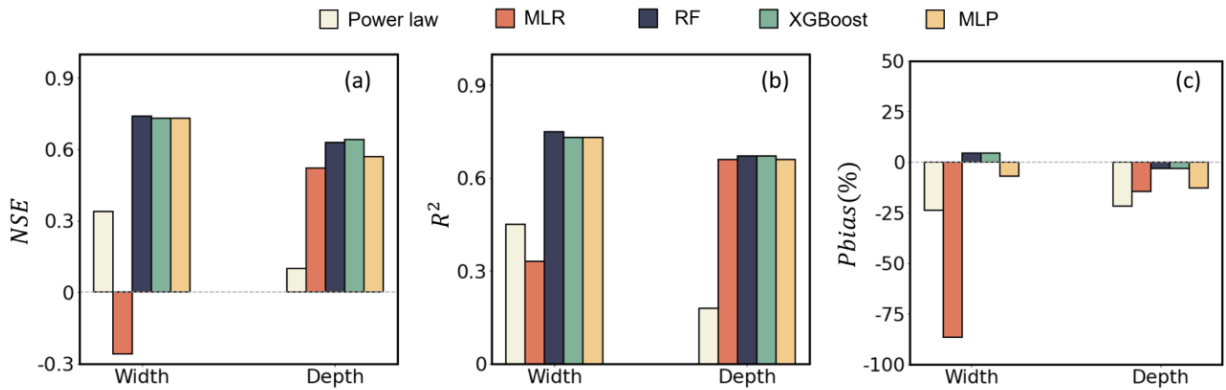

**Figure 4.** Model performance results for the river width and depth predictions: (a) NSE, (b) R$^2$, and (c) PBIAS

### 3.3 Model Performance Across Stream Orders

Interestingly, while the overall model performance is very similar for all three of the machine-learning models, the performance varies across different stream orders (**Figure 5**, **SI Tables S13, S14**). For example, using the traditional power-law models for river width and depth results in an approximately 24% underprediction of median width values and a 22% underprediction of depth values (**Figure 4c**). However, if we compare performance across stream orders, we see notable variations in performance (**Figure 5a, 5b**). The best performance for width is obtained for the 4th- and 5th-order streams (PBIAS = 8.9% and 4.6%), while we see much poorer performance for the higher-order streams, with underpredictions of width typically 50% or greater (**Figure 5(a)**). For depth, the power-law relationship provides the greatest overpredictions for 5th-order streams (median PBIAS =16.0%), but underprediction for both 1st-order and the highest-order streams (median PBIAS = - 22.5% (1st-order) and -54.23%(9th-order) ) (**Figure 5b**).

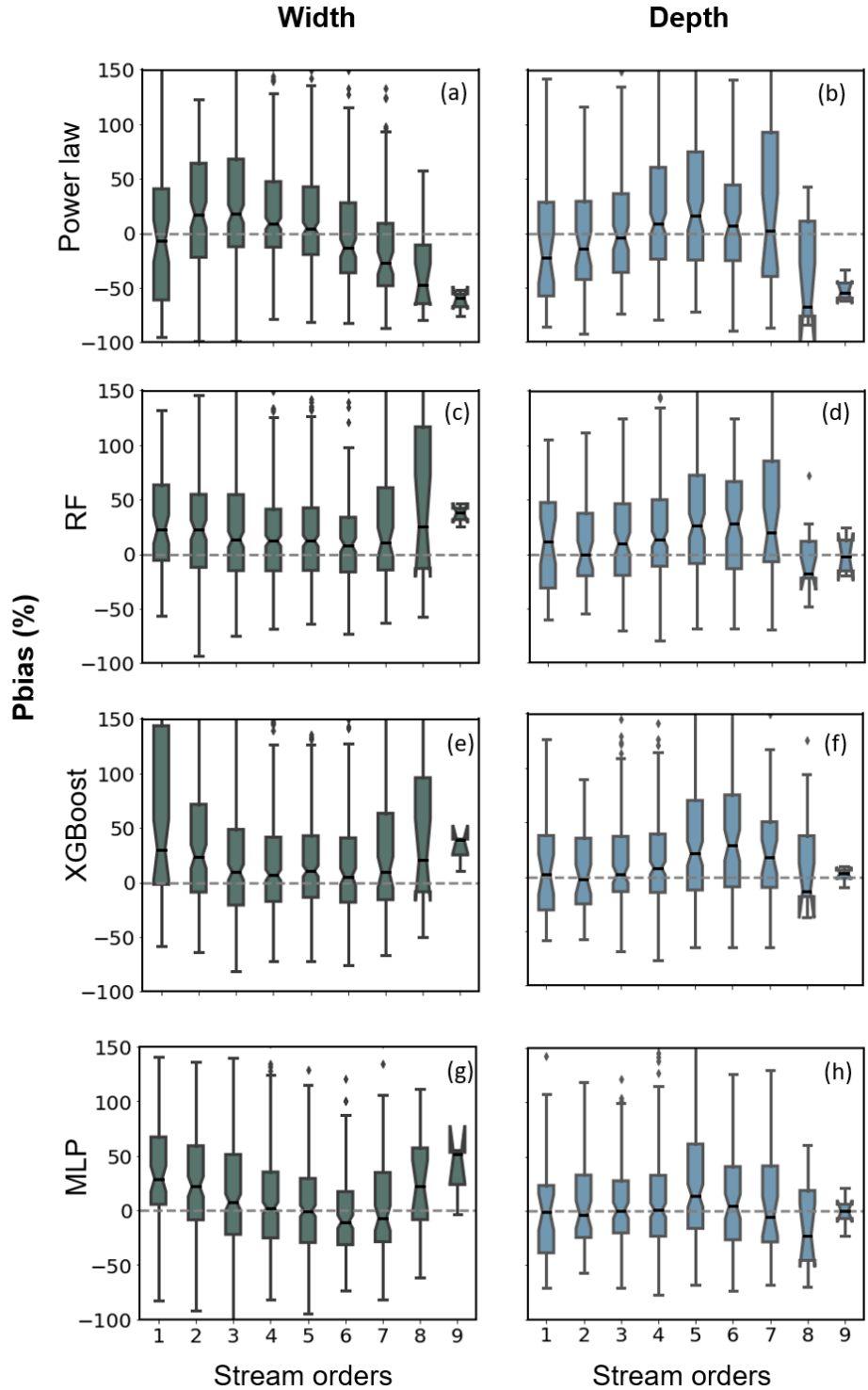

**Figure 5.** Variations in model performance as a function of stream order using the traditional power-law approach (a, b) as well as the random forest (c, d), XGBoost (e, f), and neural network (MLP) (g, h) models based on PBIAS values.

For the machine-learning approaches, we see much better overall results, with overall under- and over-prediction of 7% or less for width, and 13% or less for depth (**Fig.4c, SI TableS11, SI Table S12**). However, we also see great variability in performance across the stream orders. First, focusing only on width, while the random forest model performs relatively consistently across the stream orders, there is some variability, with the poorest performance and the greatest overprediction occurring for the lowest and highest stream orders (**Fig. 5c**). For XGBoost and the MLP neural network models, however, there is a more pronounced U-shaped behavior in the performance results for median width across stream orders (**Fig. 5e, 5g**). For $4^{th}$- and $5^{th}$-order streams, the models behave comparably in their width predictions, with the neural network demonstrating the best results. However, for $1^{st}$-order streams, the random forest performance is better than that of both the XGBoost and neural network models, as reflected by not only the median PBIAS values (median $PBIAS_{RF}$ = 22.6%, median $PBIAS_{XGB}$ = 30.0%, median $PBIAS_{MLP}$ =28.9%) but also by the narrower spread of PBIAS ($IQR_{RF}$ =69.1%, $IQR_{XGB}$ = 142.0%, $IQR_{MLP}$ = 61.81%). Variability in model performance across stream orders can also be seen for the depth predictions (**Fig. 5d, 5f, 5h**).

One hypothesis regarding the variability in model performance across stream orders might be that performance is driven by data availability. Specifically, the training data used to develop all three of the machine-learning models used in the current study disproportionately represents $4^{th}$- and $5^{th}$-order streams (**Fig. 6a, 6b**), thus potentially explaining the better performance for mid-level streams seen in the width predictions. This relationship between data availability and model performance does not hold, however, for the depth predictions, as a greater tendency toward overprediction of depth is seen for $5^{th}$-order streams in the random forest, XGBoost, and neural network models, despite the greater data availability (**Fig. 5d, 5f, 5h**). Regardless of the cause, this variability in results across stream orders, and the variability in results across models, suggests that care must be taken in model selection for predictions of geometry, particularly for the lowest-order streams, which make up the majority of waterways across the US (**Figure 6c**).

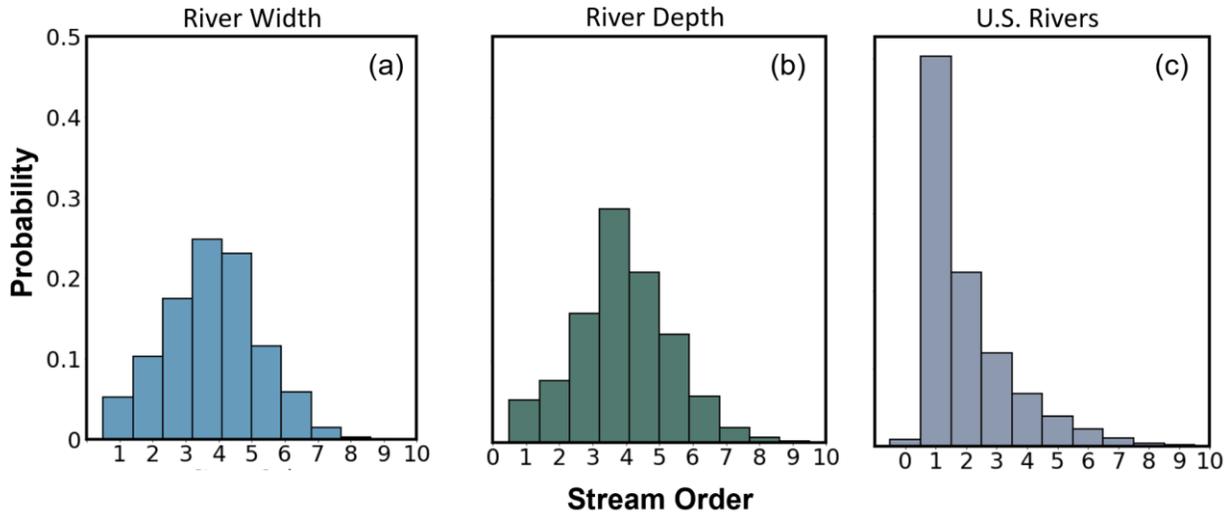

**Figure 6.** Stream order distributions. Panel (a) shows the distribution of stream orders for all available width data. Panel (b) shows the distribution for all available depth data. Panel (c) shows the distribution of stream orders for all rivers and streams across the contiguous U.S. Note that the sampling data represents an over-sampling of middle-order streams, while first-order streams are under-sampled.

### 3.4    Spatial Variations in Model Performance

We also see variations in model performance across regions and landscape types. As an example, In **Figure 7**, we provide mapped NSE values for the regionalized power-law models for width and depth as well as for our random forest, XGBoost, and MLP models at the scale of the Zone II North American Ecoregions (ERs) (US EPA, 2015). While the biggest takeaway from the figure is the greater general performance of the machine-learning models over the traditional power-law predictions, we can also see regional differences in performance with the individual models. As an example, for width, the random forest, XGBoost, and MLP models all perform very well in the Southeastern U.S. plains (ER 8.3) (NSE >0.84). In contrast, in the Appalachian region of the Eastern Temperate Forests (ER 8.4), where overall accuracy is lower for all of the machine-learning models, the random forest model (NSE = 0.68) performed better than the XGBoost model (NSE = 0.59). Further west, the MLP width model (NSE = 0.70) greatly outperforms the XGBoost model (NSE = 0.39) in the South Central semi-arid prairies (ER 9.4), while the XGBoost model (NSE = 0.86) outperforms the MLP model (NSE = 0.72) in the Upper Gila Mountains area of the Temperate Sierras (ER 13.1). Finally, in the mediterranean region of California (Ecoregion 11.1), XGBoost (NSE = 0.81) and MLP models (NSE = 0.86) both demonstrate stronger performance than the random forest model (NSE = 0.73).

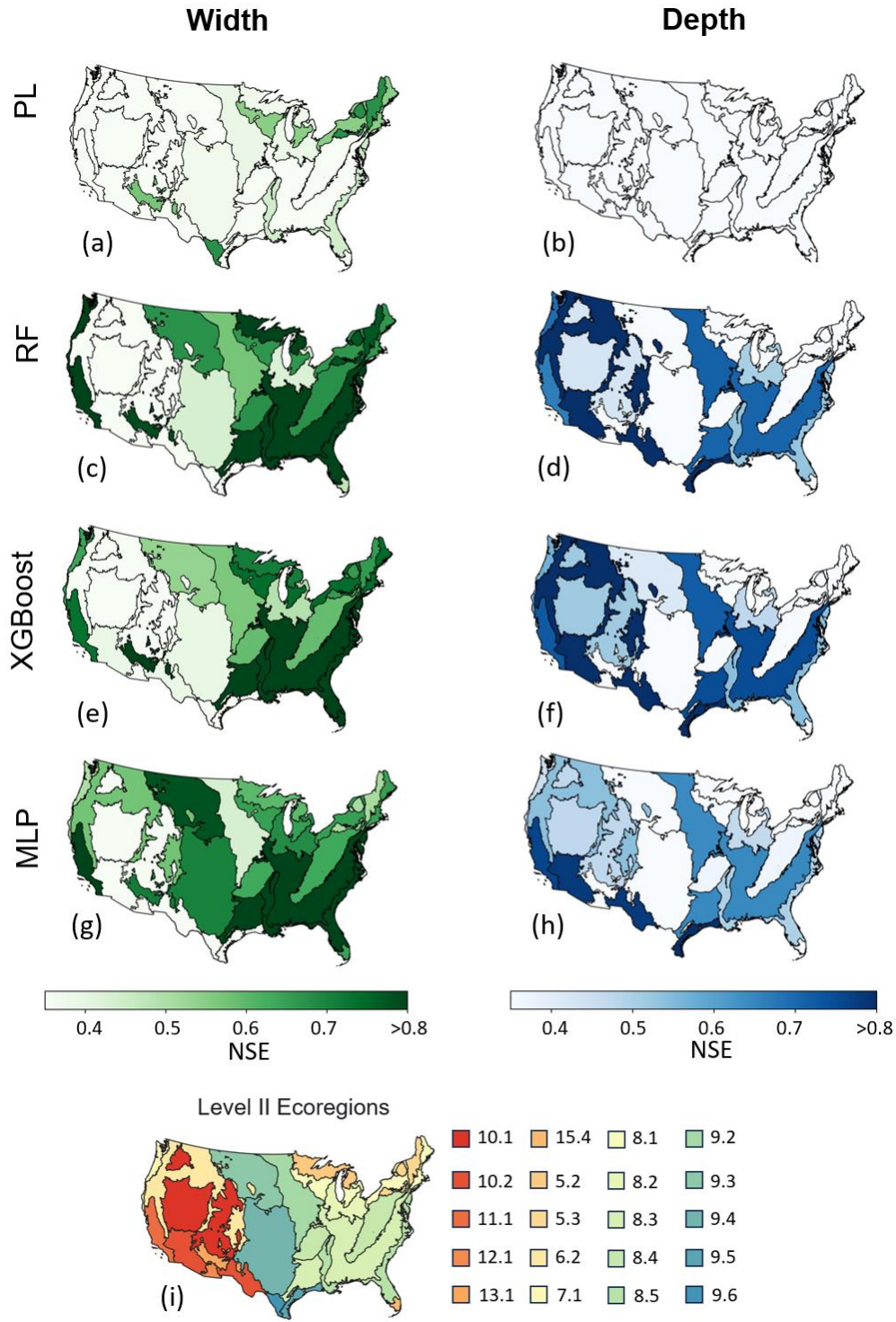

**Figure 7.** Spatially varying model performance to predict channel width and depth. Figure panels show Nash-Sutcliffe Efficiency values for the four different modeling approaches: (a, b) power law; (c, d) Random Forest; (e, f) XGBoost; (g, h) MLP; (i) Zone II North American Ecoregions (ERs).

To further demonstrate not only the spatial variability of model performance but also the differences in that spatial variance among the different models, we provide a map of differences in model performance, as indicated by differences in NSE values (ΔNSE), for both width and depth (**Figure 8**). Specifically, as an example, we are comparing here the differences in model performance between the random forest (RF) and the multilayer perceptron (MLP) models. In the figure, dark blue represents better model performance for the MLP model and red better performance for the random forest model, with white or light colors suggesting little or no difference in model performance. Here, we see that, for width, the MLP model provides much better performance than the random forest model in the Northwestern Forested Mountain zones (ER 6.2), the Cold Desert zones (ER 10.1), the South Central semi-arid prairies (ER 9.4), and the Everglades region of Florida (ER 15.4). In contrast, the random forest width model strongly outperforms the MLP width model in Marine West Coast Forest region (ER 7.1), the southwestern Warm Desert region (ER 10.2) and the Mixed Wood Shield (ER 5.2) and Atlantic Highland (ER 5.3) zones of the Northern Forest. For depth, however, the results are quite different. As an example, for depth, the random forest model outperforms the MLP model in the Everglades (ER 15.4) and the Atlantic Highlands (ER 5.3), which is a complete reversal of the performance results for channel depth. For a spatial comparison of model performance for the XGBoost and MLP models, see **SI Figure S6**. The detailed model performance across different ERs can be found in **SI Table S15** for river width and **SI Table S16** for river depth.

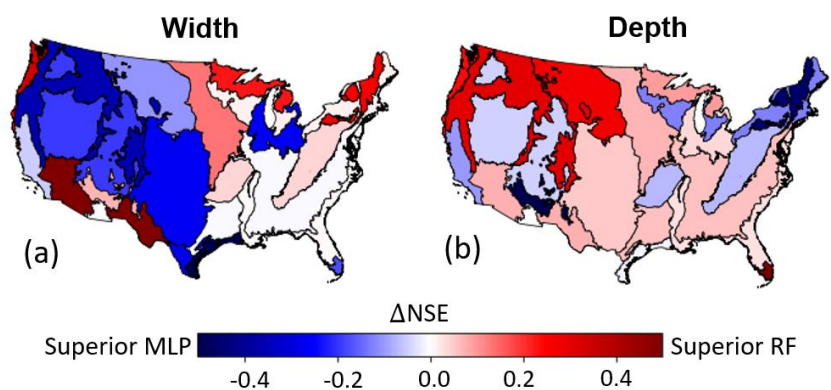

**Figure 8.** Differences in performance between the random forest and multilayer perceptron models. The maps show the difference in NSE values, ΔNSE, between the two different models for (a) width and (b) depth. Darker blue represents superior performance for the multilayer perceptron (MLP) models, darker red represents superior performance for the random forest (RF) models, and white or light colors suggest little or no difference in model performance.

In general, these results suggest that in large models such as the National Water Model, which rely on predictions of river geometry across large spatial scales for real-time predictions of river discharge, it may be beneficial to employ a multi-model approach that allows for selective use of modeled results based on regional differences in model performance. The same conclusion could also be drawn when making width and depth predictions across stream orders, with different models being selected for use with different stream orders.

## 3.5 Stream Reach and Metrics Dataset (STREAM-geo)

Predicted median river width and depth results from the machine-learning models have been made available as the U.S. Stream Reach and Metrics dataset (STREAM-geo) at the figshare data repository (available upon paper acceptance). In the published dataset, we provide results from the models with the highest overall model performance for width (XGBoost) and depth (Random Forest). STREAM-geo provides river geometry data for nearly 2.7 million NHDPlus stream reaches across the contiguous US. This newly assembled dataset provides, to our knowledge, the best available predictions of river width and river depth for all NHDPlus stream reaches. In **Figure 9,** we show the mapped STREAM-geo river width and depth data (**Fig. 9a, 9d**) along with statistical summaries of width and depth across the contiguous U.S. (**Fig. 9b, 9e**) as well as by stream order (**Fig. 9c, 9f**).

In addition to river depth and width, STREAM-geo also provides both reach-scale predictions of the river width-to-depth ratio (**Fig. 10a, 10b, 10c**) and HUC8-scale estimates of the percent river and stream surface area (%RSSA) (**Fig. 10d, 10e**) under the median flow condition. As seen in **Fig. 10b**, the width-to-depth ratio varies greatly across stream orders, ranging from a median value of 4.7 for first-order streams to 54 for tenth-order streams. This is presumably because the downstream reaches are more depositional, so they tend to have flat and wider geometries. However, the ratios plateau above $8^{th}$ order streams. Providing a more complete quantification of river width-to-depth ratios, a key parameter in fluvial geomorphology, is likely to facilitate future exploration of river characteristics, behaviors, and the processes influencing their formation.

The %RSSA is one of the principal parameters used in large-scale evaluations of river-atmosphere biogeochemical processes and thermal flux (Allen & Pavelsky, 2018). Our data demonstrates that, across the contiguous U.S., an area of nearly 50,000 km² is covered by streams and rivers, constituting 0.62% of the continental United States. These results are in alignment with a global %RSSA estimation of $0.58 \pm 0.06\%$ of Earth's non-glaciated land surface based on a global database of planform river hydromorphology and a statistical approach (Allen & Pavelsky, 2018). The Eastern half of the CONUS generally has much larger %RSSA than the West (except near the northwestern coast). Large %RSSA values are mostly found for regions with large rivers, but some upstream areas could have larger %RSSA than downstream areas, e.g., where Ohio and Missouri flow into the Mississippi. Understanding %RSSA and its distribution across different watersheds and regions will assist in the identification of biogeochemical exchange hotspots with the atmosphere.

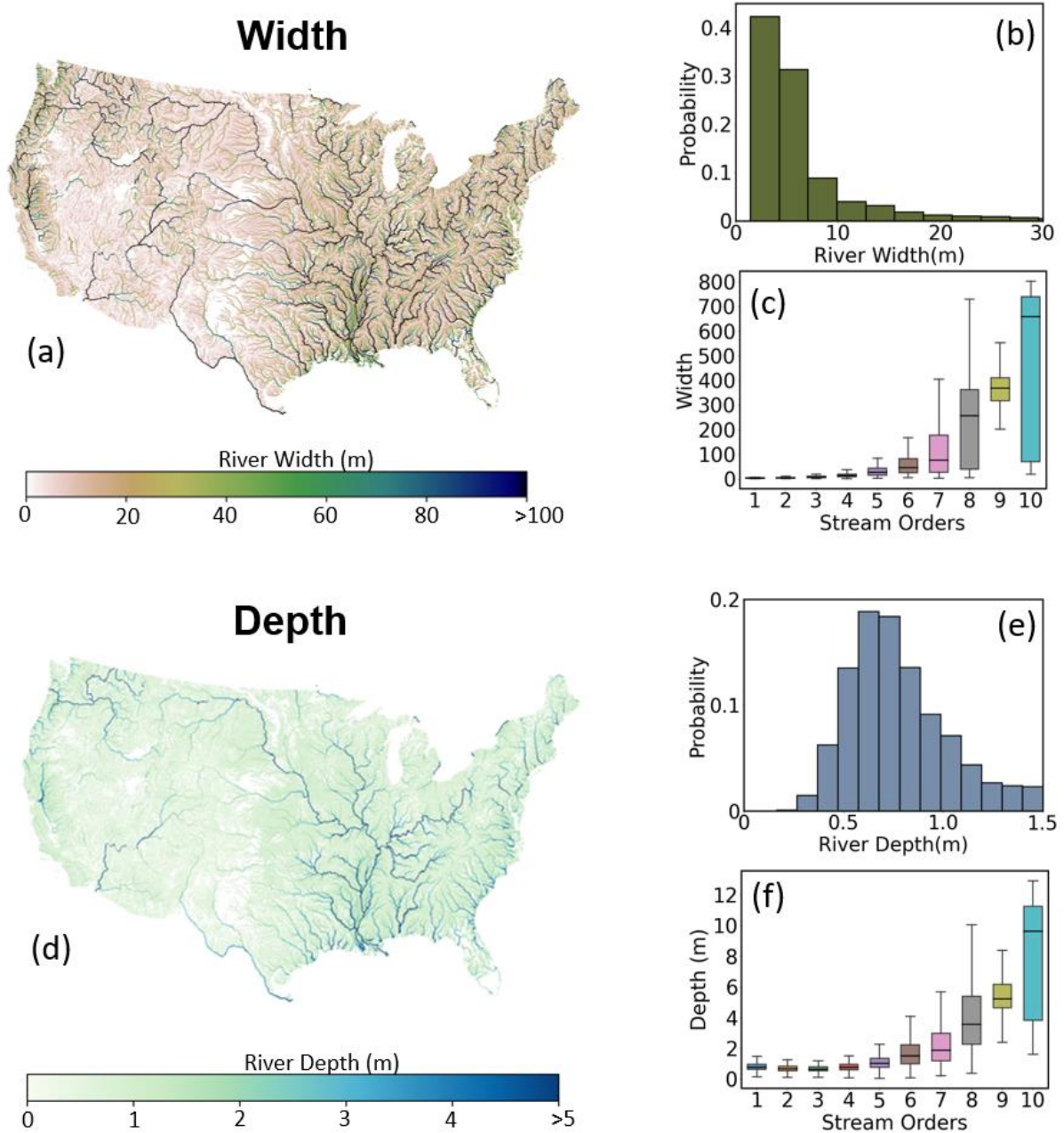

**Figure 9.** STREAM-geo Modeled stream geometry results for the contiguous US. The U.S. maps show the median river (a) width and (d) depth at the scale of individual stream reaches. (b, e) Mean river width and depth distributions across the CONUS. (c, f) River width and depth across stream orders.

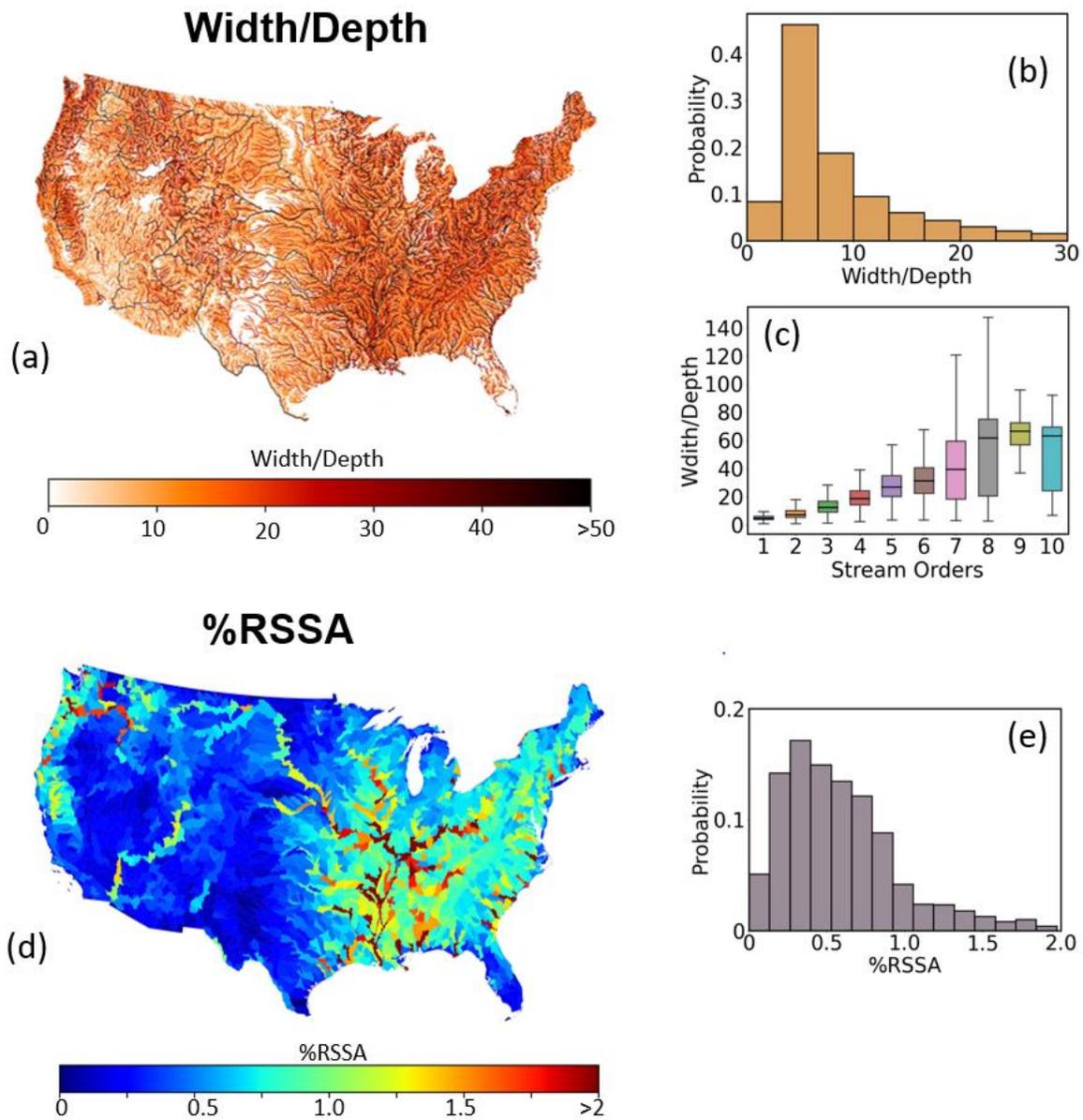

**Figure 10.** STREAM-geo river width/depth ratios and percent river surface area (%RSSA) values. The U.S. maps show (a) the river width/depth ratios at the scale of individual NHDPlus stream reaches and (d) %RSSA values at the scale of HUC8 watersheds. (b) River width/depth distributions across the CONUS. (c) River width/depth ratios across stream orders. (e) %RSSA distributions across HUC8 watersheds.

## 4.0 Conclusions

In the current study, we have demonstrated significant advancements in at-a-station hydraulic geometry (AHG) predictions from both spatial and temporal perspectives on a continental scale, meticulously compiled from extensive field measurements and machine learning techniques. All machine-learning models provided exceptional evaluation metrics that outperformed state-of–the-art regional (physiographic) hydraulic geometry equations, developed across hydrological landscape regions (HLRs), which have been widely used in large-scale hydrology models for flood prediction and inundation mapping, including the National Water Model.

It is important to note here that the high level of performance with the current modeling approaches was achieved using only publicly available land surface information, without the need for detailed subsurface altimetry and satellite information. The recently launched Surface Water and Ocean Topography (SWOT) mission, is expected to provide new measurements of river height, slope, and width for rivers greater than 100 m in width, with the potential to obtain data even for rivers as small as 50 m in width. While the SWOT mission data will provide a tremendous advance in river geometry data, it must also be noted that our results show that the vast majority (> 96.9%) of US stream reaches are narrower than that 50-m threshold. In the current study, however, our models were trained based on extensive data from first-order headwater streams, and thus our results can meaningfully supplement the new SWOT observations to fill the gap for smaller streams.

Our models also clearly demonstrate the utility of using multi-model machine-learning approaches to characterize the complex relationships between catchment attributes, riverbank characteristics, discharge, and river hydrogeomorphological characteristics. Our findings show that the performance of different modeling approaches differs not only by stream order but also by geographical region, even when using the same set of underlying predictors. These results clearly show that large-scale models like the National Water Model, which are tasked with real-time discharge predictions across vast geographical expanses, stand to gain significantly from a multi-model strategy. Further development of such multi-model strategies can not only refine predictions but also constitute a progressive step toward creating adaptive, multi-faceted analytical frameworks in the advancement of water system science.

Finally, the results of the present work have resulted in the development of STREAM-geo, a comprehensive river geometry database for the continental United States. This dataset includes NHDPlus reach-scale predictions of river width, depth, width-to-depth ratios, and stream surface area for 2.7 million reaches, thus providing a groundbreaking new resource to promote new insights into stream and river systems to support hydrology, ecology, and river management. These now publicly available predictions of river geometry parameters can be further applied to both data-driven and physical-based models for simulations and predictions related to flood management and enhanced understanding of erosion, sediment transport, aquatic ecosystems, and beyond.

## Acknowledgements

The authors declare that there are no conflicts of interest in publishing the current work. The work was funded by the National Weather Service and the Consortium of Universities for the Advancement of Hydrologic Science, Inc. (CUAHSI), during the 2022 NOAA-CUAHSI National Water Center Innovators Program. The two science advisors (Dr. Chaopeng Shen and Dr. Sagy Cohen) were funded by the Cooperative Institute for Research to Operations in Hydrology (CIROH), award numbers A22-0307-S003 and A22-0305, respectively.

## Open Research

The STREAM-Geo dataset produced as a result of this research will be publicly available upon paper acceptance. A complete list of the model outputs included in STREAM-geo can be found in Appendix A. All input data used in the models is publicly available and is described in the Methods and listed in Table S2 in the Supplemental Information.

## Appendix A

STREAM-geo includes two CSV files (STREAM-geo_River_Geometry.csv and STREAM-geo_River_%RSSA_HUC8.csv), and two GeoPackage files (STREAM-geo_River_Geometry.gpkg and STREAM-geo_River_%RSSA_HUC8.gpkg).

Table A1. Column name, unit, and content of STREAM-geo_River_Geometry

| Field Name | Descriptions | Unit | Source |
|---|---|---|---|
| COMID | Common identifier of an NHDFlowline feature | | NHDPlus |
| StreamOrde | Modified Strahler Stream Order | | NHDPlus |
| w_m | River reach width | Meters ($m$) | Chang et al |
| d_m | River reach depth | Meters ($m$) | Chang et al |
| rssa_m2 | River and stream surface area | Squared meters ($m^2$) | Chang et al |
| q_va | Annual mean discharge | Cubic feet per second ($cfs$) | Chang et al |
| LENGTHKM | River reach length | Kilometers ($km$) | NHDPlus |

Table A2. Column name, unit, and content of STREAM-geo_River_%RSSA_HUC8

| Field Name | Descriptions | Unit | Source |
|:---:|:---:|:---:|:---:|
| Huc8 | Common identifier of HUC8 sub-watershed | | USGS HUC8 |
| Rssa m2 | River and stream surface area within each HUC8 sub-watershed | Squared meters ($m^2$) | Chang et al |
| Rssa% | River and stream surface area percentage | % | Chang et al |

# Supplement

# The geometry of flow: Advancing predictions of river geometry with multi-model machine learning

**Authors:** S.Y. Chang[1,2], Zahra Ghahremani[3], Laura Manuel[4], Mohammad Erfani[5], Chaopeng Shen[6], Sagy Cohen[7], Kimberly Van Meter[1,3], Jennifer L Pierce[3], Ehab A Meselhe[4], Erfan Goharian[5]

[1]*Department of Geography, Pennsylvania State University, 302 Walker Building, University Park, PA 16803*

[2]*Earth and Environmental Systems Institute, 2217 Earth-Engineering Sciences Building*

*University Park, PA 16802-6813*

[3]*Department of Geoscience, Boise State University, Environmental Research Building 1160, Boise, ID 83725*

[4]*Department of River-Coastal Science and Engineering, Tulane University, 6823 St. Charles Avenue*

*New Orleans, LA 70118*

[5]*Department of Civil and Environmental Engineering, Univerity of South Carolina, 300 Main St, Room C206, Columbia, SC 29208*

[6]*Department of Civil and Environmental Engineering Engineering,  Pennsylvania State University, 231C Sackett Bulding, University Park, PA 16803*

[7]*Department of Geography, University of Alabama, Shelby Hall 2019-E, Tuscaloosa, AL, 35487*

**Corresponding authors' Email:** cshen@engr.psu.edu

sagy.cohen@ua.edu

vanmeterkvm@psu.edu

# Table of Contents

# S1.0  River Geometry Dataset

**Table S1**. Post-processed HYDRoSWOT river geometry observations and statistics

| Target Variable | River Width $\underline{w}$ | River Depth $\underline{d}$ |
|---|---|---|
| # Observations | 172491 | 51039 |
| # Streams | 5269 | 2678 |
| Mean (m) | 44.71 | 1.52 |
| Std(m) | 72.14 | 1.42 |
| Min (m) | 0.27 | 0.2 |
| 25% (m) | 13.51 | 0.72 |
| 50% (m) | 24.23 | 1.11 |
| 75% (m) | 45.42 | 1.77 |
| Max (m) | 1091.2 | 16.87 |

**Table S2.** Detailed information of the predictor variable collections: variable name, descriptions, resolution, processing, and sources.

| Variable name | Variable name | Descriptions | Processes/ Resolution | Data sources |
|---|---|---|---|---|
| Q | q_va [cfs] | Discharge (cfs) | Point measurement | Canova et al., 2016 |
| Dam | ACC_NDAMS2010 | Accumulated number of dams built on or before 2010 based on total upstream accumulation | Polyline based | Schwarz et al., 2018 |
| Aridity | AI1 | Global averaged aridity | Resolution of | Zomer et al., |

| | | index 1970–2000 for the 1970–2000 period | 30 arc-seconds, within 500m site buffer | 2022 |
|---|---|---|---|---|
| Population | CAT_POPDENS10 [persons/ km2] | Population density, persons per square kilometer from 2010 Census block level data | Individual reach catchments, Polyline based | (Schwarz et al., 2018) |
| D50 | D50[mm] | Median bed-material sediment particle size in mm | Polyline based | (Schwarz et al., 2018) |
| Elevation | MINELEVSMO [cm] | Minimum elevation (smoothed) in centimeters | Polyline based | NHDPlus Version 2.0 |
| EVI-summer | EVI_JAS_2012 | Averaged enhanced Vegetation Index for 2012 summer (July, August, and September) | Individual reach catchments scale, Polyline based | (Schwarz et al., 2018) |
| EVI-winter | EVI_JFM_2012 | Averaged enhanced Vegetation Index for 2012 winter (January, February, and March) | Individual reach catchments scale, Polyline based | (Schwarz et al., 2018) |
| Ag% | NLCD_agriculture_16 [%] | Percent of agricultural land use from the new generation of the 2016 National | Individual reach catchments scale, Polyline based | (Schwarz et al., 2018) |
| Dev% | NLCD_developed_16[%] | Percent of developed land use from the new generation of the 2016 National | Individual reach catchments scale, Polyline based | Schwarz et al., 2018 |
| Forest% | NLCD_forest_16 [%] | Percent of forest land use from the new generation of the 2016 National | Individual reach catchments scale, Polyline based | Schwarz et al., 2018 |
| Slope | SLOPE | Slope of flowline based on smoothed elevations | Polyline based | NHDPlus Version 2.0 |
| Order | StreamOrde | Modified Strahler Stream Order | Polyline based | NHDPlus Version 2.0 |
| Area | TotDASqKM [km2] | Total upstream catchment area (km2) | Polyline based | NHDPlus Version 2.0 |
| Clay% | Percent_Clay1km _0_100cm[%] | Mean clay content in the top 100cm of soil | Resolution of 30 m, within 500m site buffer | gSSURGO |
| Sand% | percent_sand1km _0_100cm[%] | Mean sand content in the top 100cm of soil | Resolution of 30 m, within 500m site buffer | gSSURGO |
| Silt% | Percent_Silt_1km 100cmfinal[%] | Mean silt content in the top 100cm of soil | Resolution of 30 m, within 500m site buffer | gSSURGO |

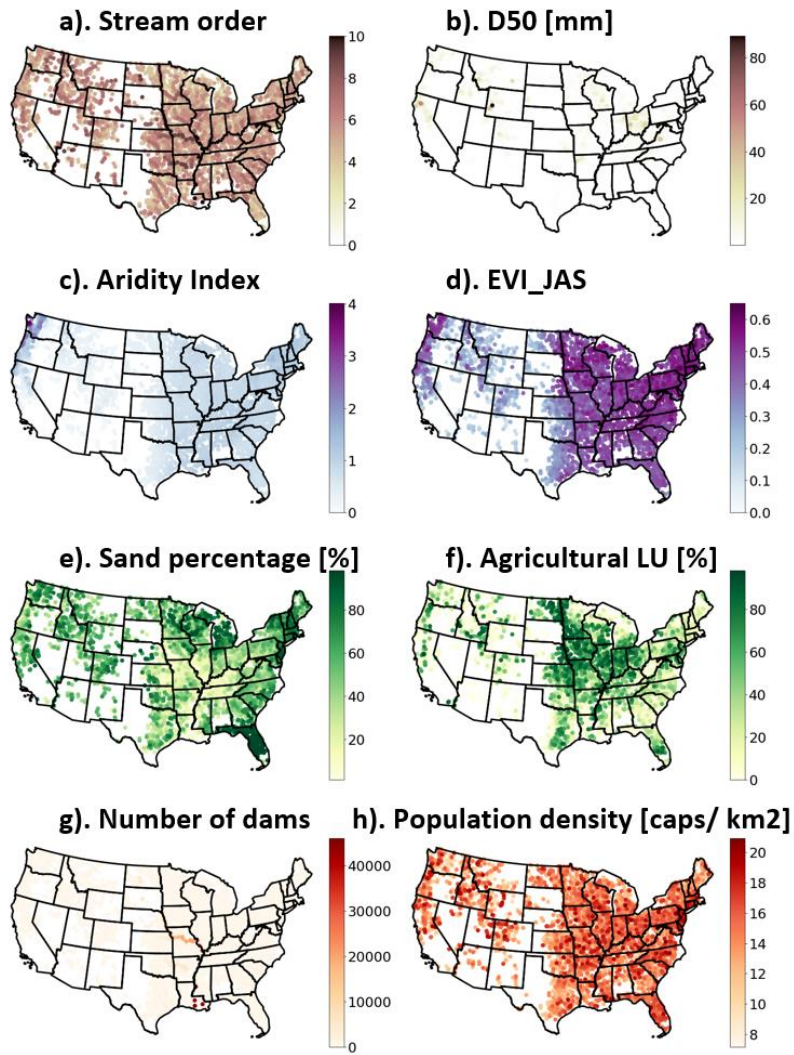

**Figure S1.** The eight selected catchment and river attributes of HYDRoSWOT gauges. (a) Stream order: the level of branching in a river system in geomorphology and hydrology; (b). D50: 50% percent of the total bedding particles; (c).Aridity index: an indicator of long-term climatic aridity; (d). EVI_JAS: 2012 summer enhanced vegetation index; (e). Sand percentage (f). Agricultural land use percentage; (g). The number of dams upstream; (h). population density.

# S2.0  Model Development

**Table S3** Calibrated hyperparameter values for random forest to predict _w_ and _d_

| Hyperparameter Searching Space | | River Width $\underline{w}$ | River Depth $\underline{d}$ |
|---|---|---|---|
| Tree Depth | range(5,15,2) | 11 | 5 |
| Minimum number of samples at a leaf node | range(3,8,1) | 3 | 3 |
| Minimum number of samples in a node | [2, 4, 8, 10] | 2 | 2 |
| Number of Trees | range(30,100,10) | 90 | 30 |

**Table S4** Calibrated hyperparameter values for the XGBoost to predict _w_ and  _d_

| Hyperparameter Searching Space | | River Width $\underline{w}$ | River Depth $\underline{d}$ |
|---|---|---|---|
| colsample_bytree | [ 0.3, 0.4, 0.5 , 0.7 ] | 0.7 | 0.7 |
| Learning rate | Eta = [0.05, 0.10, 0.15, 0.20, 0.25, 0.30 ] | 0.1 | 0.1 |
| Minimum loss reduction | Gamma = [ 0.0, 0.1, 0.2 , 0.3, 0.4 ] | 0 | 0.1 |
| Maximum depth of a tree | [ 3, 4, 5, 6, 8, 10, 12, 15] | 5 | 4 |

| min_child_weight | [ 1, 3, 5, 7 ] | 5 | 1 |

**Table S5** Calibrated hyperparameter values for MLP to predict _w_ and _d_

| Hyperparameter Searching Space | | River Width _w_ | River Depth _d_ |
|---|---|---|---|
| #hidden layers | m=i{1,2,3,4, 5} | 2 | 5 |
| #Number of neurons within each hidden layer | nm=i{range(50,330,20)} | 440, 480 | 60,200,140, 100, 100 |
| Dropout ratio for each hidden layer | dm=i{0, 0.1, 0.2, 0.3} | 0.1, 0 | 0, 0, 0.2, 0.2, 0.2 |
| # Epochs | k=i{range(30,100,10)} | 120 | 20 |
| Learning rate | l=i{range(1e-5,1e-3)} in logspace | 0.000119378 | 0.000150131 |
| Batch size | b=i{range(30,160, 20)} | 60 | 20 |

**Table S6.** Scaling methods for input and output variables to prepare for MLPs

| Variables | Transformation Functions | Mathematical equations |
|---|---|---|
| q_va, D50, SLOPE | Power Law | $x_t = x^{1/3}$ |
| MINELEVSMO,percent_sand1km_0_100cm,AI1,NLCD_agriculture_16,TotDASqKM,ACC_NDAMS2010,Percent_Clay1km_0_100cm,Percent_Silt_1km100cmfinal,EVI_JFM_2012 | Normalization | $x_t = \dfrac{x - \underline{x}}{\sigma}$ |
| All | Max-min normalization | $x_t = \dfrac{x - x_{min}}{x_{max} - x_{min}}$ |

# S3.0 Linear Regression for Variable Importances

## S3.1 Linear regression for median width

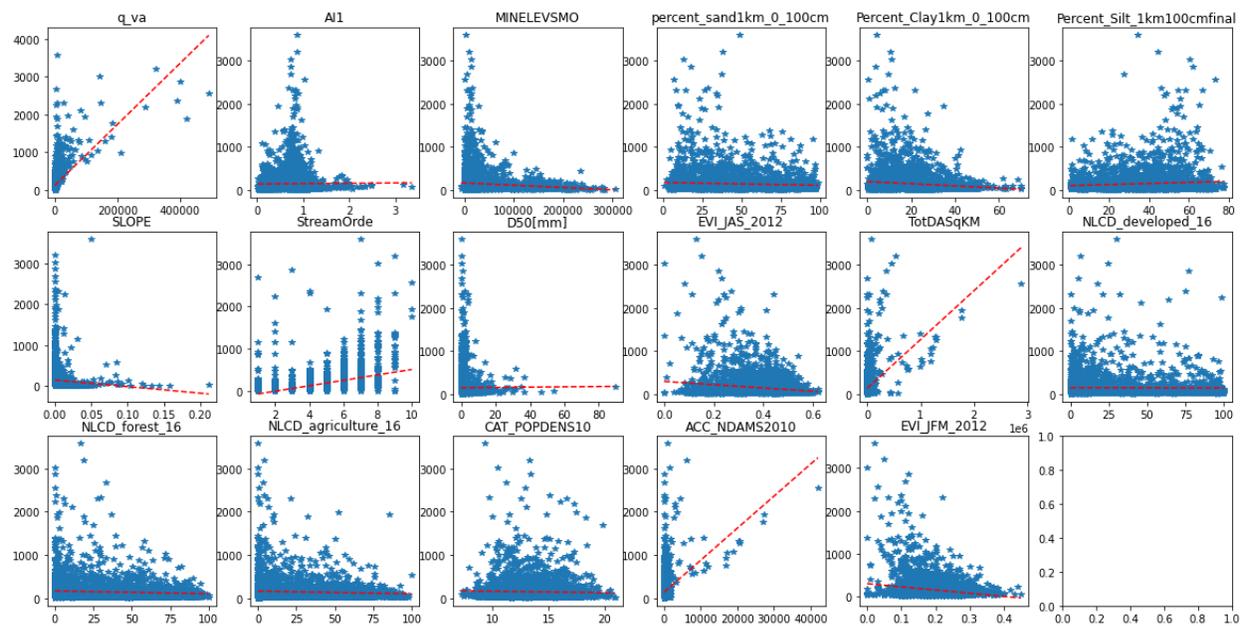

**Figure S2**. Relationships between the 17 attributes and median river width through linear regression.

**Table 7.** The explained covariance, slope, intercept, and P-value from simple linear regressions for median river width

| Variable Name | r2 | slope | intercept | p-value |
|---|---|---|---|---|
| Q | 0.36496 | 0.00807 | 125.26840 | 0.00000 |

| | | | |
|---|---|---|---|
| Aridity | 0.00011 | 8.10746 | 140.31060 | 0.52445 |
| Elevation | 0.00982 | -0.00052 | 163.51480 | 0.00000 |
| Sand% | 0.00390 | -0.63701 | 173.46540 | 0.00010 |
| Clay% | 0.01370 | -2.53480 | 196.45630 | 0.00000 |
| Silt% | 0.00747 | 1.25315 | 101.42530 | 0.00000 |
| Slope | 0.00396 | -1598.85000 | 151.86840 | 0.00009 |
| Order | 0.19500 | 64.06750 | -119.95500 | 0.00000 |
| D50 | 0.00002 | 0.35065 | 145.86220 | 0.75802 |
| EVI-summar | 0.03116 | -379.00400 | 292.38820 | 0.00000 |
| A | 0.20395 | 0.00113 | 131.52980 | 0.00000 |
| Dev% | 0.00006 | -0.06433 | 147.87400 | 0.63865 |
| Forest% | 0.00543 | -0.64405 | 164.47610 | 0.00000 |
| Ag% | 0.00462 | -0.62681 | 160.68010 | 0.00002 |
| Population | 0.00083 | -3.14276 | 188.47130 | 0.07332 |
| Dam | 0.16592 | 0.07422 | 135.15960 | 0.00000 |
| EVI-winter | 0.04056 | -745.95100 | 292.80210 | 0.00000 |

## S3.2 Linear regression for median depth

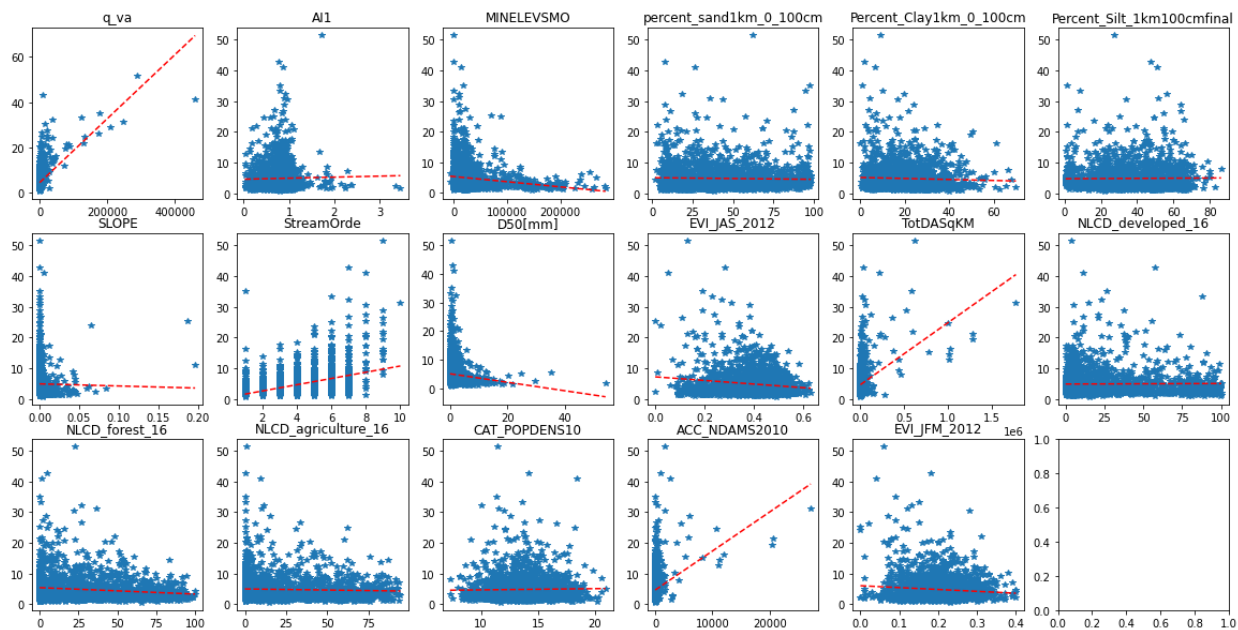

**Figure S3**. Relationships between the 17 attributes and median river depth through linear regression.

**Table 8** The explained covariance, slope, intercept, and P-value from simple linear regressions for median river depth

| Variable Name | r2 | slope | intercept | p-value |
|---|---|---|---|---|
| Q | 0.439577 | 0.000121 | 4.100822 | 0 |
| Aridity | 0.001779 | 0.599702 | 4.0382 | 0.30286 |
| Elevation | 0.028314 | -1.45E-05 | 5.039589 | 0 |
| Sand% | 0.000809 | -0.00537 | 4.703946 | 0.16563 |
| Clay% | 0.01138 | -0.04452 | 5.370973 | 0.05038 |
| Silt% | 0.001164 | 0.009029 | 4.147802 | 0.58394 |
| Slope | 0.008309 | -43.2935 | 4.637279 | 0.52327 |
| Order | 0.172497 | 1.100156 | -0.10497 | 0 |
| D50 | 0.004409 | -0.07469 | 4.599749 | 0.00001 |
| EVI-summar | 0.033691 | -7.04192 | 7.197045 | 0 |
| A | 0.256632 | 1.43E-05 | 4.198118 | 0 |
| Dev% | 9.19E-05 | -0.00151 | 4.516157 | 0.63905 |
| Forest% | 0.016198 | -0.02019 | 5.049514 | 0 |
| Ag% | 0.002916 | -0.00865 | 4.691368 | 0.04997 |
| Population | 0.002105 | 0.089477 | 3.29421 | 0.35229 |
| Dam | 0.242436 | 0.000955 | 4.226415 | 0 |
| EVI-winter | 0.010643 | -7.07959 | 5.835029 | 0.00002 |

# S3.3 Linear regression for median width (log-log)

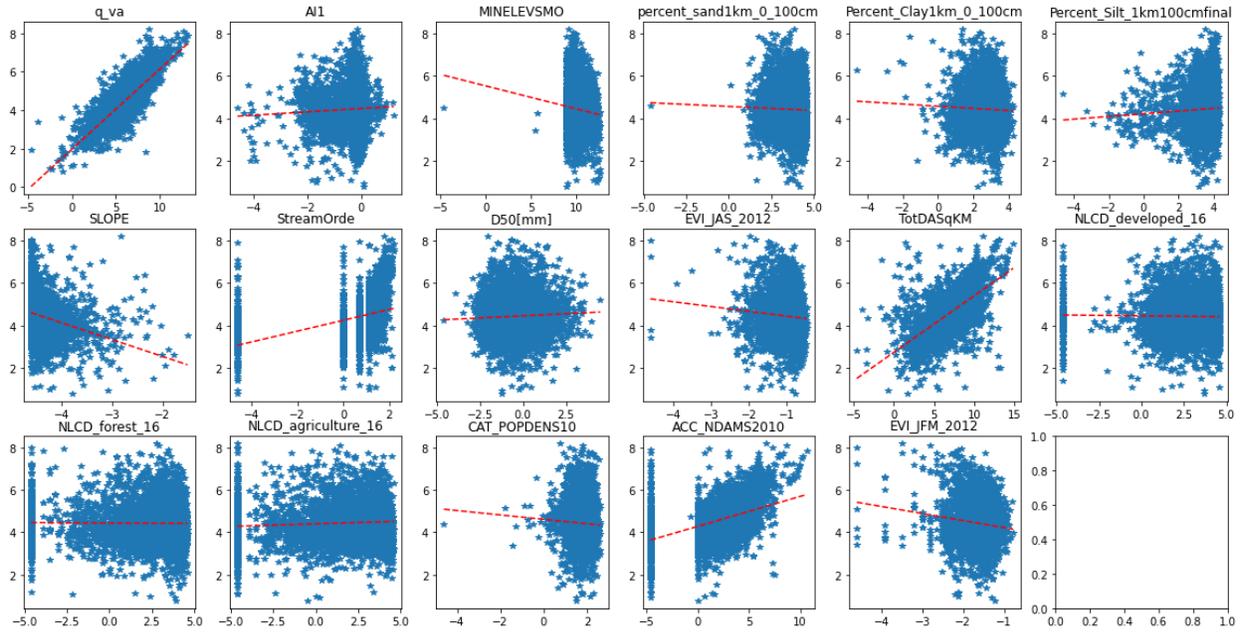

**Figure S4**. Relationships between the 17 attributes and median river width through log-log linear regressions.

**Table 9.** The explained covariance, slope, intercept, and P-value from log-log linear regressions for median river width

| Variable Name | r2 | a | b | p-value |
|---|---|---|---|---|
| Q | 0.606 | 7.088 | 0.419 | 0 |
| Aridity | 0.001 | 86.964 | 0.079 | 0.001958 |
| Elevation | 0.006 | 252.933 | -0.108 | 0 |
| Sand% | 0.002 | 95.911 | -0.038 | 0.084551 |
| Clay% | 0.013 | 96.776 | -0.051 | 0.009846 |
| Silt% | 0.005 | 67.887 | 0.064 | 0.000352 |
| Slope | 0.024 | 2.572 | -0.794 | 0 |
| Order | 0.075 | 69.229 | 0.256 | 0 |
| D50 | 0.001 | 85.394 | 0.04 | 0.003227 |
| EVI-summar | 0.04 | 67.244 | -0.226 | 0 |
| A | 0.358 | 15.196 | 0.267 | 0 |
| Dev% | 0 | 85.511 | -0.008 | 0.32137 |
| Forest% | 0.003 | 84.61 | -0.004 | 0.517205 |
| Ag% | 0 | 81.359 | 0.025 | 0.000003 |
| Population | 0.001 | 100.584 | -0.103 | 0.005319 |
| Dam | 0.207 | 72.04 | 0.141 | 0 |
| EVI-winter | 0.068 | 48.732 | -0.332 | 0 |

## S3.4 Linear regression for median depth (log-log)

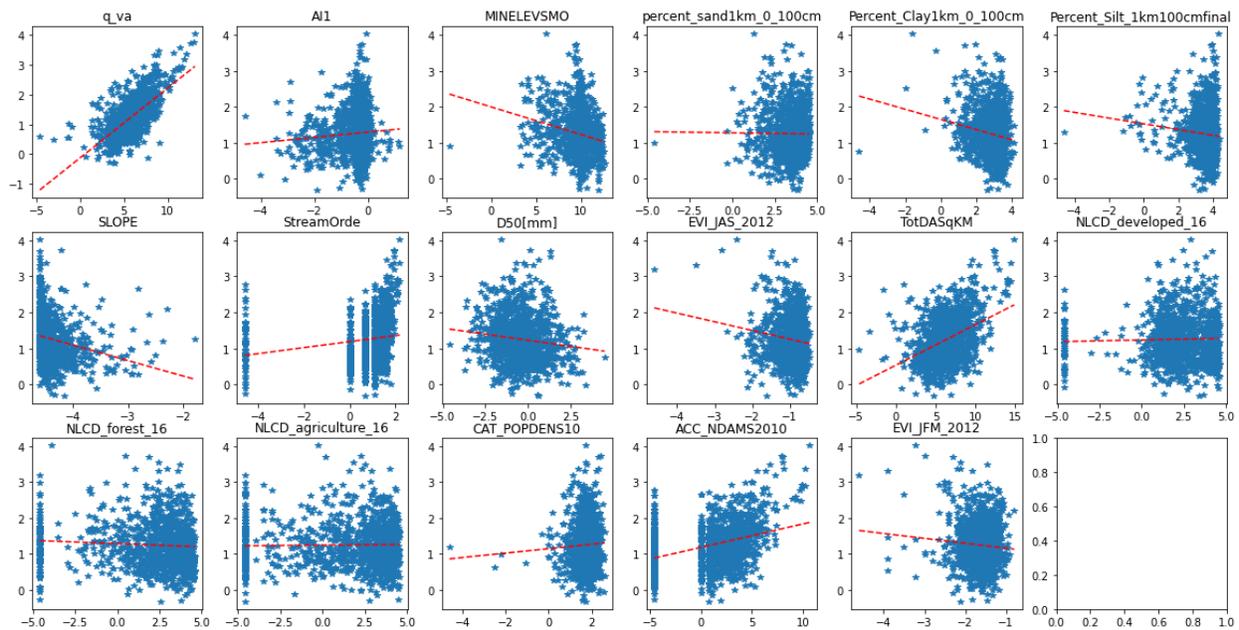

**Figure S5.** Relationships between the 17 attributes and median river depth through log-log linear regressions.

**Table10.** The explained covariance, slope, intercept, and P-value from log-log linear regressions for median river depth

| Variable Name | r2 | a | b | p-value |
|---|---|---|---|---|
| Q | 0.569 | 0.886 | 0.234 | 0 |
| Aridity | 0.006 | 3.648 | 0.074 | 0.04958 |
| Elevation | 0.025 | 7.366 | -0.077 | 0 |
| Sand% | 0.004 | 3.577 | -0.007 | 0.01466 |
| Clay% | 0.038 | 5.191 | -0.141 | 0.00151 |
| Silt% | 0.001 | 4.591 | -0.08 | 0.00117 |
| Slope | 0.034 | 0.532 | -0.43 | 0 |
| Order | 0.033 | 3.274 | 0.083 | 0 |
| D50 | 0.005 | 3.398 | -0.068 | 0 |
| EVI-summer | 0.063 | 2.758 | -0.241 | 0.00031 |
| A | 0.25 | 1.698 | 0.113 | 0 |
| Dev% | 0.002 | 3.428 | 0.009 | 0.0721 |
| Forest% | 0.005 | 3.625 | -0.018 | 0.00001 |
| Ag% | 0 | 3.478 | 0.004 | 0.10406 |
| Population | 0.003 | 3.151 | 0.062 | 0.33824 |
| Dam | 0.158 | 3.271 | 0.065 | 0 |
| EVI-winter | 0.038 | 2.787 | -0.136 | 0.00695 |

# S4.0  Model performance

## S4.1 Multiple-linear Regression Equations

Building upon the extensive literature that has employed log-log linear regression, multiple linear regression models were also developed for this study using JMP statistical software and were optimized for each target dependent variable under median-flow conditions (channel width and depth), utilizing our newly organized training dataset, encompassing not only watershed area and discharge but also numerous other catchment and bank properties. This can be used to test the optimal model formula for river geometry simulation. The median width and median depth datasets were utilized and all variables were log-transformed, given the high degree of skewness in most variables. All variables presented in **Table S2.** were used to initiate the model. A forward stepwise regression technique was then used to determine the significance of each variable to the model performance. Variables were initially removed based on p-value criteria, with p-values >0.05 being considered not significant. Next, variables were removed one by one, beginning with the least significant, so as to not deteriorate the model prediction statistic.

$$\underline{w} = 10^{0.853} \times Q^{0.321} \times AI^{0.1976} \times Emin^{-0.0381} \times Slope^{-0.0265} \times A^{0.125} \qquad \text{SE1}$$

$$\underline{d} = 10^{0.43} \times Q^{0.238} \times Emin^{-0.061} \times Sand\%^{-0.114} \times Silt\%^{-0.112} \times Slope^{-0.035} \times D50^{-0.0035} \qquad \text{SE2}$$

## S4.2 Model performance metrics

**Table S11** Model performance for the estimation of median river width

| $\underline{w}$ | Overall Matrix | | | |
|---|---|---|---|---|
| | $R^2$ | NSE | PBIAS | RMSE (m) |
| Power-Law | 0.45 | 0.34 | -23.87% | 52.76 |
| MLR | 0.33 | -0.26 | -86.61% | 73.04 |
| RF | 0.75 | 0.74 | 4.61% | 32.94 |
| XGBoost | 0.73 | 0.73 | 4.41% | 34 |
| MLP | 0.73 | 0.73 | -7.00% | 33.74 |

**Table S12** Model performance for the estimation of median river depth

| _d_ | Overall Matrix | | | |
|---|---|---|---|---|
| | R2 | NSE | PBIAS | RMSE |
| Power-Law | 0.18 | 0.1 | -21.92% | 1.62 |
| MLR | 0.66 | 0.52 | -14.68% | 1.19 |
| RF | 0.67 | 0.63 | -3.23% | 1.05 |
| XGBoost | 0.67 | 0.64 | -2.98% | 1.03 |
| MLP | 0.66 | 0.57 | -12.89% | 1.12 |

## S4.3 Model Performance difference across different stream orders

**Table S13** The median pbias of River Width models across different stream orders

| Stream Order | River Width (_w_) | | | |
|---|---|---|---|---|
| | Power-law | RF | XGBoost | MLP |
| 1 | -7.21% | 22.64% | 30.04% | 28.90% |
| 2 | 17.09% | 22.25% | 23.83% | 21.80% |
| 3 | 17.75% | 13.46% | 9.39% | 7.40% |
| 4 | 8.91% | 12.16% | 6.67% | 2.21% |
| 5 | 4.56% | 12.13% | 10.46% | -0.54% |
| 6 | -13.46% | 8.28% | 5.46% | -10.51% |
| 7 | -27.21% | 10.88% | 9.36% | -7.23% |
| 8 | -47.09% | 25.56% | 20.44% | 22.52% |
| 9 | -59.15% | 37.87% | 39.53% | 51.30% |
| 10 | -67.55% | -0.53% | -2.26% | -10.01% |

**Table S14** The midian pbias of River depth models across different stream orders

| Stream Order | River Depth($\underline{\boldsymbol{d}}$) | | | |
|:---:|:---:|:---:|:---:|:---:|
| | Power-law | RF | XGBoost | MLP |
| 1 | -22.47% | 11.82% | 2.32%% | -1.11% |
| 2 | -13.93% | -0.75% | -2.43% | -3.29% |
| 3 | -3.98%% | 9.86% | 2.21% | 0.49% |
| 4 | 8.45% | 13.82% | 8.02% | 0.70% |
| 5 | 16.03% | 26.10% | 21.45% | 14.00% |
| 6 | 7.10% | 27.90% | 29.14% | 4.24% |
| 7 | 2.02% | 19.43% | 18.28% | -5.06% |
| 8 | -67.32% | -17.96% | -12.93% | -22.68% |
| 9 | -54.27% | -2.25% | 3.31% | -0.22% |
| 10 | -83.39% | -39.98% | -37.02% | -61.42% |

## S4.4 Model Performance difference across different regions

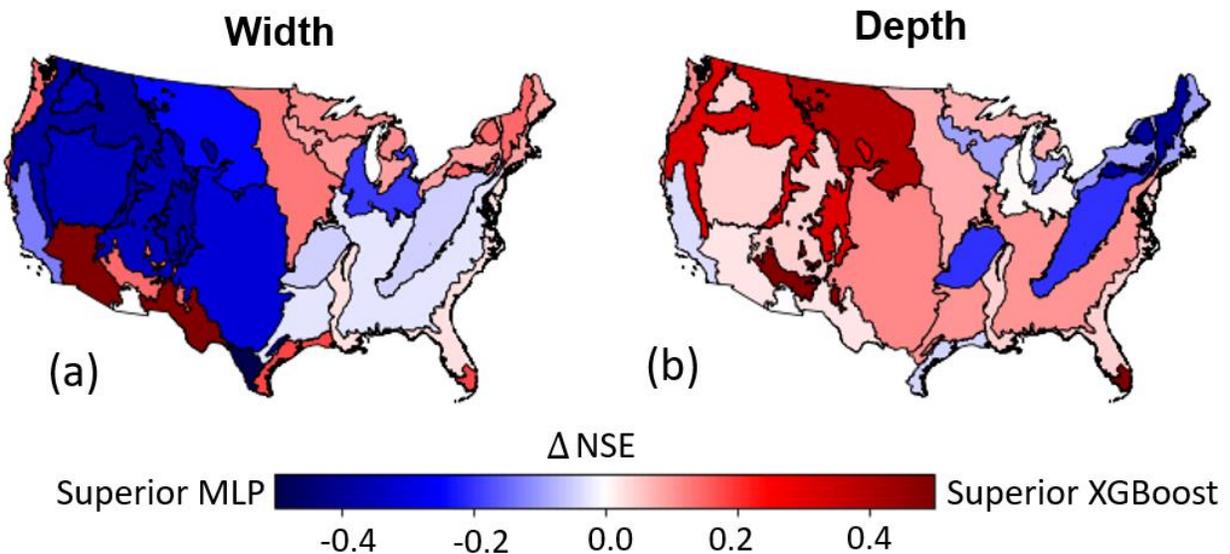

**Figure S6.** Differences in performance between the XGBoost and multilayer perceptron models.

**Table S15** Model performances for river width in different Ecoregions level II.

| Ecoregion | Number of records | Power-law NSE | RF NSE | XGBoost NSE | MLP NSE |
|---|---|---|---|---|---|
| 5.2 | 39 | 0.373722 | 0.828053 | 0.711967 | 0.600372 |
| 5.3 | 38 | 0.675442 | 0.770539 | 0.657267 | 0.510708 |
| 6.2 | 92 | 0.211513 | 0.215929 | 0.194154 | 0.57505 |
| 7.1 | 35 | 0.269142 | 0.839211 | 0.642263 | 0.504346 |
| 8.1 | 148 | 0.547405 | 0.664935 | 0.741274 | 0.650491 |
| 8.2 | 81 | 0.385692 | 0.424413 | 0.490961 | 0.68405 |
| 8.3 | 277 | 0.215382 | 0.856726 | 0.837981 | 0.863748 |
| 8.4 | 154 | 0.269801 | 0.675758 | 0.589767 | 0.633559 |
| 8.5 | 100 | 0.441442 | 0.859443 | 0.885023 | 0.855226 |
| 9.2 | 136 | 0.390447 | 0.571689 | 0.565545 | 0.43585 |
| 9.3 | 26 | 0.210693 | 0.674175 | 0.533769 | 0.77727 |
| 9.4 | 86 | 0.374167 | 0.43336 | 0.394736 | 0.702713 |
| 9.5 | 7 | -0.4154 | -2.76839 | -0.43417 | -0.61568 |
| 9.6 | 2 | 0.676825 | -4.45055 | -13.2729 | -4.17437 |
| 10.1 | 62 | 0.217913 | 0.162005 | 0.013945 | 0.354245 |
| 10.2 | 8 | -0.1947 | 0.268672 | 0.390276 | -0.42349 |
| 11.1 | 15 | 0.234693 | 0.811942 | 0.732307 | 0.860067 |
| 12.1 | 2 | -19.0926 | -1.54769 | -1.55892 | -3.03748 |
| 13.1 | 7 | 0.570341 | 0.786811 | 0.859504 | 0.713456 |
| 15.4 | 4 | -0.41786 | 0.471631 | 0.825601 | 0.644117 |

**Table S16** Model performances for river depth in different Ecoregions level II.

| Ecoregion | Number of records | Power-law NSE | RF NSE | XGBoost NSE | MLP NSE |
|---|---|---|---|---|---|
| 5.2 | 23 | -0.60981 | 0.003847 | -0.01519 | -0.09262 |
| 5.3 | 24 | 0.13942 | -0.4852 | -0.4219 | -0.03063 |
| 6.2 | 43 | 0.029132 | 0.840577 | 0.855538 | 0.538189 |
| 7.1 | 16 | -0.17393 | 0.692513 | 0.56678 | 0.451995 |
| 8.1 | 56 | -0.08673 | 0.2159 | 0.248234 | 0.338511 |
| 8.2 | 36 | 0.114817 | 0.502987 | 0.472809 | 0.467797 |
| 8.3 | 136 | 0.12709 | 0.707124 | 0.7493 | 0.64511 |
| 8.4 | 69 | 0.157925 | 0.301822 | 0.168029 | 0.36949 |
| 8.5 | 83 | -0.01378 | 0.527911 | 0.540333 | 0.496936 |
| 9.2 | 59 | 0.004188 | 0.713043 | 0.7218 | 0.641074 |
| 9.3 | 8 | -0.13862 | 0.284584 | 0.413671 | 0.014021 |
| 9.4 | 37 | -0.36401 | 0.214916 | 0.276859 | 0.161768 |
| 9.5 | 5 | -0.81876 | 0.858508 | 0.826865 | 0.867549 |
| 9.6 | 1 | | | | |
| 10.1 | 39 | 0.114817 | 0.426369 | 0.517656 | 0.472887 |
| 10.2 | 4 | 0.057805 | 0.856554 | 0.805924 | 0.78027 |
| 11.1 | 9 | -0.04551 | 0.662996 | 0.727178 | 0.76205 |
| 13.1 | 2 | -2.25109 | -5.9946 | -0.67112 | -3.69625 |
| 15.4 | 4 | -2.1447 | -1.081 | -0.84546 | -1.68157 |